\documentclass[apj]{emulateapj}
\usepackage{comment}

\shorttitle{FIR-RADIO CORRELATION AT HIGH-REDSHIFTS}
\shortauthors{MURPHY}

\begin{document}
\title{The Far-Infrared--Radio Correlation at High Redshifts: Physical Considerations and Prospects for the Square Kilometer Array}

\author{Eric J. Murphy}
\affil{{\it Spitzer} Science Center, California Institute of Technology, MC 314-6, Pasadena CA, 91125; emurphy@ipac.caltech.edu} 
\slugcomment{Draft version 2.1  September 29, 2009}
\journalinfo{Accepted to \apj, September 29, 2009}

\begin{abstract}
I present a predictive analysis for the behavior of the far-infrared (FIR)--radio correlation as a function of redshift in light of the deep radio continuum surveys which may become possible using the square kilometer array (SKA).  
To keep a fixed ratio between the FIR and predominantly non-thermal radio continuum emission of a normal star-forming galaxy, whose cosmic-ray (CR) electrons typically lose most of their energy to synchrotron radiation and Inverse Compton (IC) scattering, requires a nearly constant ratio between galaxy magnetic field and radiation field energy densities.  
While the additional term of IC losses off of the cosmic microwave background (CMB) is negligible in the local Universe, the rapid increase in the strength of the CMB energy density (i.e. $\sim(1+z)^{4})$ suggests that evolution in the FIR-radio correlation should occur with infrared (IR;~$8-1000~\micron$)/radio ratios increasing with redshift.  
This signature should be especially apparent once beyond $z\sim3$ where the magnetic field of a normal star-forming galaxy must be  $\sim$50~$\mu$G to save the FIR-radio correlation.  
At present, observations do not show such a trend with redshift; 
$z\sim6$ radio-quiet quasars appear to lie on the local FIR-radio correlation while a sample of $z\sim4.4$ and $z\sim2.2$ submillimeter galaxies (SMGs) exhibit ratios that are a factor of $\sim$2.5 {\it below} the canonical value.  
I also derive a 5$\sigma$ point-source sensitivity goal of $\approx$20~nJy (i.e. $\sigma_{\rm RMS} \sim 4$~nJy) requiring that the SKA specified be $A_{\rm eff}/T_{\rm sys}\approx  15000$~m$^{2}$~K$^{-1}$; 
achieving this sensitivity should enable the detection of galaxies forming stars at a rate of $\ga25~M_{\sun}~{\rm yr}^{-1}$, such as typical luminous infrared galaxies (i.e. $L_{\rm IR} \ga 10^{11}~L_{\sun}$), at all redshifts if present.   
By taking advantage of the fact that the non-thermal component of a galaxy's radio continuum emission will be quickly suppressed by IC losses off of the CMB, leaving only the thermal (free-free) component, I argue that deep radio continuum surveys at frequencies $\ga$10~GHz may prove to be the best probe for characterizing the high-$z$ star formation history of the Universe unbiased by dust.  
\end{abstract}

\section{Introduction}
Radio continuum emission from galaxies arises due to a combination of thermal and non-thermal processes primarily associated with the birth and death of young massive stars, respectively.  
The thermal (free-free) radiation of a star-forming galaxy is emitted from H{\sc ii} regions and is directly proportional to the photoionization rate of young massive stars.  
Since emission at GHz frequencies is optically thin, the thermal radio continuum emission from galaxies is a very good diagnostic of a galaxy's massive star formation rate.  

Massive ($\ga 8~M_{\sun}$) stars which dominate the Lyman continuum luminosity also end their lives as supernovae (SNe) whose remnants (SNRs) are responsible for the acceleration of cosmic-ray (CR) electrons into a galaxy's general magnetic field resulting in diffuse synchrotron emission.  
Thus, in a more complicated manner, the non-thermal radio continuum emission also traces the most recent star formation activity in a galaxy.  
Ideally, one would like to isolate the thermal component as it is a more direct measure of the most  recent massive star formation activity.  
However, at GHz frequencies, the non-thermal fraction typically dominates the total radio continuum emission \citep[i.e. $\sim$10:1 at $\sim$1~GHz;][]{cy90} making the isolation of the thermal fraction difficult. 


 These same massive stars are often the primary sources of dust heating in the interstellar medium (ISM) as their starlight is absorbed and reradiated at far-infrared (FIR) wavelengths by interstellar grains.  
 This common origin between the FIR dust emission and thermal $+$ non-thermal radio continuum emission from galaxies is thought to be the dominant physical processes driving the FIR-radio correlation on global \citep[e.g.][]{de85,gxh85,sn97,nb97,yrc01} and local \citep[e.g.][]{bg88,xu92,mh95,hoer98,hip03,ejm06,ejm08,ah06} scales.   
 While massive star formation provides a shared origin between emission at these two wavelengths, a number of physical processes must conspire to yield such a remarkably constant ratio among galaxies spanning nearly 5 orders of magnitude in luminosity  \citep[e.g.][]{yrc01}, among other properties (e.g. Hubble type, FIR color, and FIR/optical ratio).  
 For example, since the non-thermal synchrotron emissivity of a single electron is roughly proportional to the electron density times the square of the magnetic field strength, it would seem that the FIR-radio correlation should be highly dependent on the propagation of CR electrons and a galaxy's magnetic field distribution and strength.  
If magnetic fields in galaxies are built-up over time, and therefore were weaker in the past, the correlation should be different at higher redshifts.  
However, all indications suggest that the FIR-radio correlation holds out to moderate redshifts \citep[e.g.][]{mg02,grup03,pa04,df06,ejm09a,ms09} suggesting that the magnetic field strength and structure in these galaxies is similar to what is observed for galaxies in the local Universe.    
Consequently, comparing the FIR and radio emission characteristics of galaxies can illuminate properties that are typically inaccessible by other observational means.  
 
 \begin{figure*}
\plottwo{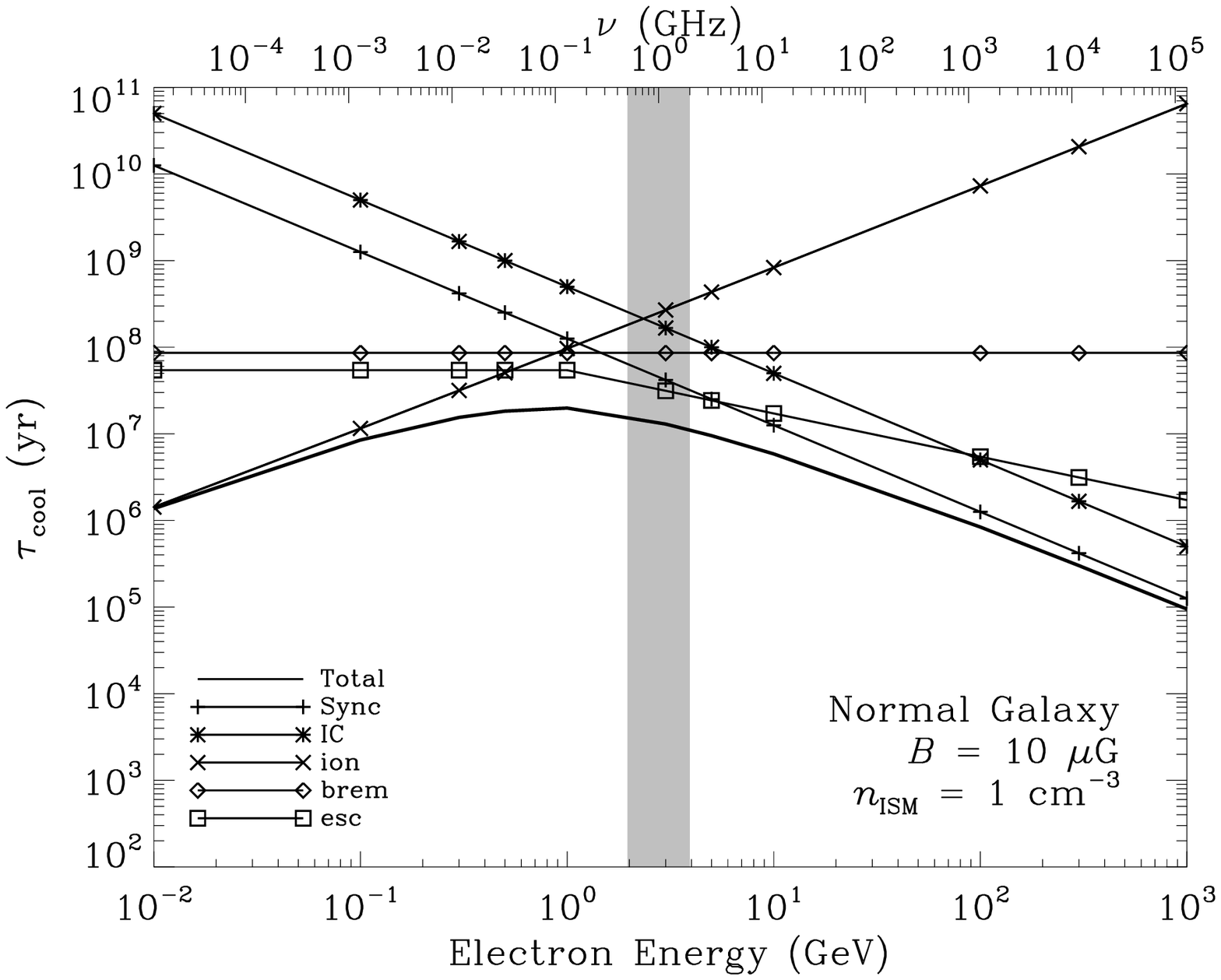}{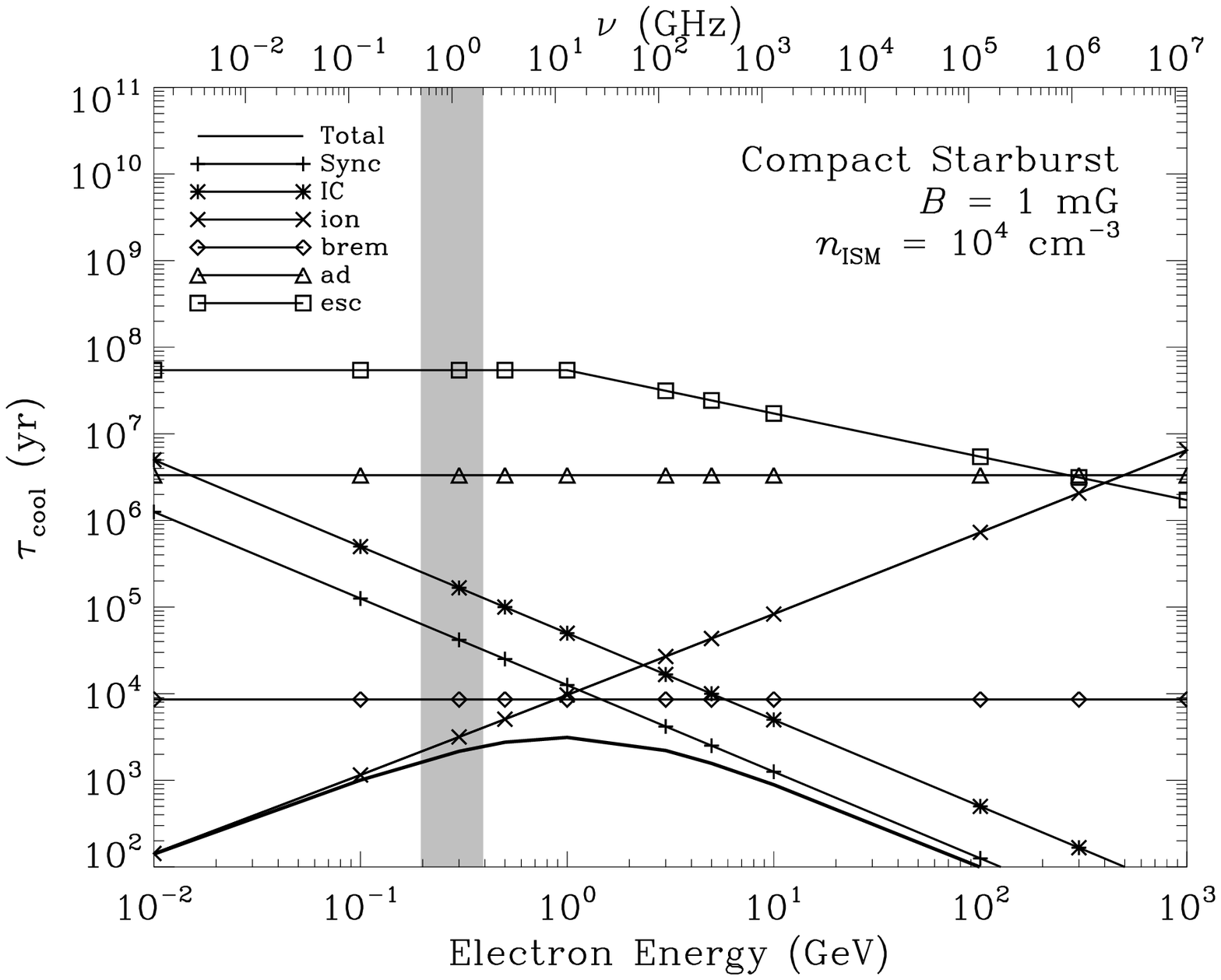}
\caption{
{\it Left:} 
The cooling timescales of CR electrons as a function of electron energy due to a number of physical processes: synchrotron, Inverse Compton (IC), ionization, and bremsstrahlung.  
For these calculations typical parameters for normal star-forming galaxies are assumed: $B=10~\mu$G, $U_{\rm rad} = 10^{-12}$~erg~cm$^{-3}$, and $n_{\rm ISM} = 1$~cm$^{-3}$. 
A timescale due to the escape of CR electrons from a galaxy, assuming random walk diffusion and an escape scale-length of $l_{\rm esc} = 3$~kpc, is also included.  
The top axis indicates the synchrotron emitting frequency of the CR electrons.  
The shaded region indicates the electron energies emitting predominantly in the $0.5-2$~GHz passband.     
{\it Right:} Same as what is plotted in the left panel, but for values that may be more typical of galaxies hosting compact starbursts, namely: $B=1$~mG and $n_{\rm ISM} = 10^{4}$~cm$^{-3}$, which assumes the usual $B \propto \sqrt{n_{\rm ISM}}$ scaling.  
The additional term from adiabatic expansion losses due to a starburst-driven galactic wind, having a velocity of $v_{w} = 300$~km~s$^{-1}$ and assuming a disk scale-height $h= 1$~kpc, are also shown.   
$U_{\rm rad}$ was also scaled up to 10$^{-8}$~erg~cm$^{-3}$.  
For normal star-forming galaxies, synchrotron and IC losses dominate CR electron energy losses, along with escape, for radiation observed at $\ga$GHz frequencies.  
However, in the case where a mG magnetic field is considered, synchrotron and IC losses no longer are the dominant cooling processes for CR electrons emitting at GHz frequencies.   
\label{fig-1}}
\end{figure*}

At present, detailed studies of the FIR-radio correlation among star-forming galaxies has been limited to only moderate redshifts at $z \la 1$.  
With the sensitivity of existing FIR and radio capabilities, studies at higher redshifts only probe the most extreme objects.   
For example, deep ($\sigma_{\rm RMS} \sim0.55~$mJy) 70$\micron$ 
 observations \citep{df06} from the the Far-Infrared Deep Extragalactic Legacy (FIDEL; PI. M. Dickinson) survey is able to probe luminous infrared galaxies (LIRGs; $10^{11} \leq L_{\rm IR} < 10^{12}~L_{\sun}$) and ultraluminous infrared galaxies  (ULIRGs;  $\geq 10^{12}~L_{\sun}$) out to redshifts of $z\sim 1$ and $z\sim2 $, respectively.   
With {\it Herschel}, surveys such as GOODS-{\it Herschel} should improve this situation by measuring the FIR properties of normal star-forming galaxies out to $z\sim 1$, and that of LIRGs and ULIRGs out to redshifts of $z\sim2$ and $z\sim4$, respectively.  
ALMA will also play a significant role by measuring the peak of the FIR spectral energy distribution (SED)  for all LIRGs at $z\ga5$.  

These populations of dusty star-forming galaxies appear to dominate the stellar mass assembly at increasing redshifts; 
the star formation rate density increases by a factor of $\sim5-10$ between $z\sim0$ and $z\sim1$,  becoming increasingly obscured (i.e. $\ga60$\%) by dust \citep{de02,ce01,el05,bm09}.  
Thus, LIRGs and ULIRGs appear to dominate the luminosity density at increasing redshifts, and optically thin measures of star formation and active galactic nuclei (AGN) activity are critical for proper quantification of stellar mass build-up over cosmic time.  
At radio wavelengths, however, detecting such high redshift galaxies remains extremely difficult.  
Even with a fully operational EVLA, IR-bright star-forming galaxies  (e.g. M~82; $L_{\rm IR} \approx 4\times10^{10}~L_{\sun}$) and moderate LIRGs  (i.e. $L_{\rm IR} \approx 3\times10^{11}~L_{\sun}$) will not be detectable beyond redshifts of $z\sim 1$ and $z\sim 2$, respectively.  

A next-generation radio facility such as the Square Kilometer Array (SKA) should easily remedy this disparity between the depth of FIR and radio continuum surveys.  
 While the SKA was initially proposed solely on the basis of H{\sc i} science, deep continuum imaging is critical for the realization of nearly all of the five established Key Science Projects (KSPs), especially  studies of ``The Origin and Evolution of Cosmic Magnetism" and ``Galaxy Evolution and Cosmology." 
 Essential to these two science goals is the proper measurement of the star formation and AGN history over cosmic time for which deep radio continuum studies may provide an excellent advantage over other wavelengths.  

In this paper I present physically motivated expectations for the behavior of the FIR-radio correlation at increasing redshift and discuss their implications.    
I also discuss how the SKA, when combined with future FIR observations to be obtained with next generation facilities both prior to, and commensurate with, final SKA science operations, will be able to tackle interesting problems associated with these KSPs.  
The paper is organized as follows: 
In $\S$2 I discuss the FIR-radio correlation and introduce the physical processes for which it depends on.  
Then, in section $\S$3, I present the results for the expected evolution in the FIR-radio correlation at increasingly high redshifts and report on the technical specifications for detecting high-$z$ galaxies in deep radio continuum surveys.    
In Section $\S$4 I discuss the physical implications of the results for characterizing the properties of galaxies at high~$z$ and compare these theoretical expectations with existing observations.  
Finally, in $\S$5, I summarize my conclusions.  


\section{The FIR-Radio Correlation: Physical Considerations}
A major result of the {\it Infrared Astronomical Satellite} \citep[IRAS;][]{gxn84} all-sky survey was the discovery of a correlation between the globally measured far infrared (FIR;~$42-122~\micron$) dust emission and the optically thin radio continuum emission of normal late-type star-forming galaxies without AGN \citep{de85,gxh85}.  
The most remarkable feature of this correlation is that it displays such little scatter (i.e. $\approx$0.26~dex) among  galaxies spanning 5 orders of magnitude in luminosity \citep{yrc01}.
While the FIR emission is due to the thermal re-radiation of interstellar starlight by dust grains, the radio emission is primarily non-thermal synchrotron emission from CR electrons that propagate in a galaxy's magnetic field after initially being accelerated by SN shocks or other processes. 
The physics that maintains a strong correlation between these two quantities over such a wide range of galaxies remains unclear, however departures from the nominal ratio can shed insight on a number of typically inaccessible galaxy properties.    

While the FIR emission 
is dependent upon the chemical make-up and  size distribution of dust grains, which is hard to decipher, a galaxy's non-thermal radio emission is relatively simple to interpret.  
What is not simple, however, is determining which energy-loss processes may dominate a galaxy's population of CR electrons, as well as the role of escape.  
CR electrons will cool via a number of mechanisms as they propagate through the ISM of galaxies.  
These energy-loss processes include synchrotron radiation and Inverse Compton (IC) scattering, as well as ionization, bremsstrahlung, and adiabatic expansion losses.  
Below I briefly describe the role that each of these energy-loss processes may play for the cases of normal star-forming galaxies and compact starbursts following \citet{ml94}.  

\subsection{Essential Physics: Normal Star-Forming Galaxies}
\begin{figure}
\plotone{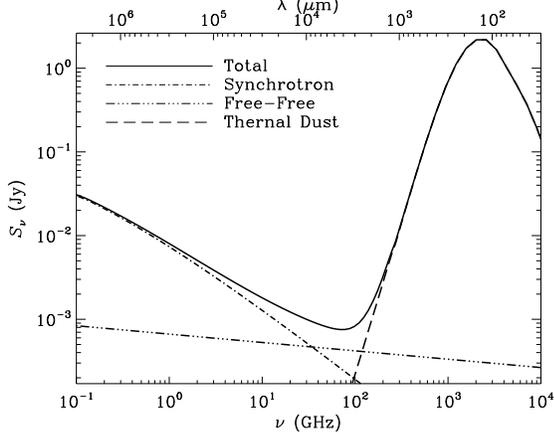}
\caption{
A model radio-FIR spectrum for a galaxy having a FIR flux of $F_{\rm FIR} = 10^{-13}$~W~m$^{-2}$.  
The FIR portion of the spectrum is given using the IR dust emission component of the \citet{dh02} SED libraries.  
The thermal radio spectrum, which goes as $S_{\nu}^{\rm T} \propto \nu^{-0.1}$, is normalized assuming that the ionizing photon rate estimated by the thermal radio continuum and IR luminosity are the same, and that the thermal fraction at 1.4~GHz is 10\%.  
Under these assumptions the nominal FIR-radio correlation (i.e. $q_{\rm IR} \approx 2.64$) is achieved.     
The non-thermal spectrum includes energy losses from all of those mechanisms considered in deriving the CR cooling timescales shown Figure \ref{fig-1} (See $\S$2.1 for details).       
\label{fig-2}}
\end{figure}

\begin{figure*}
\plotone{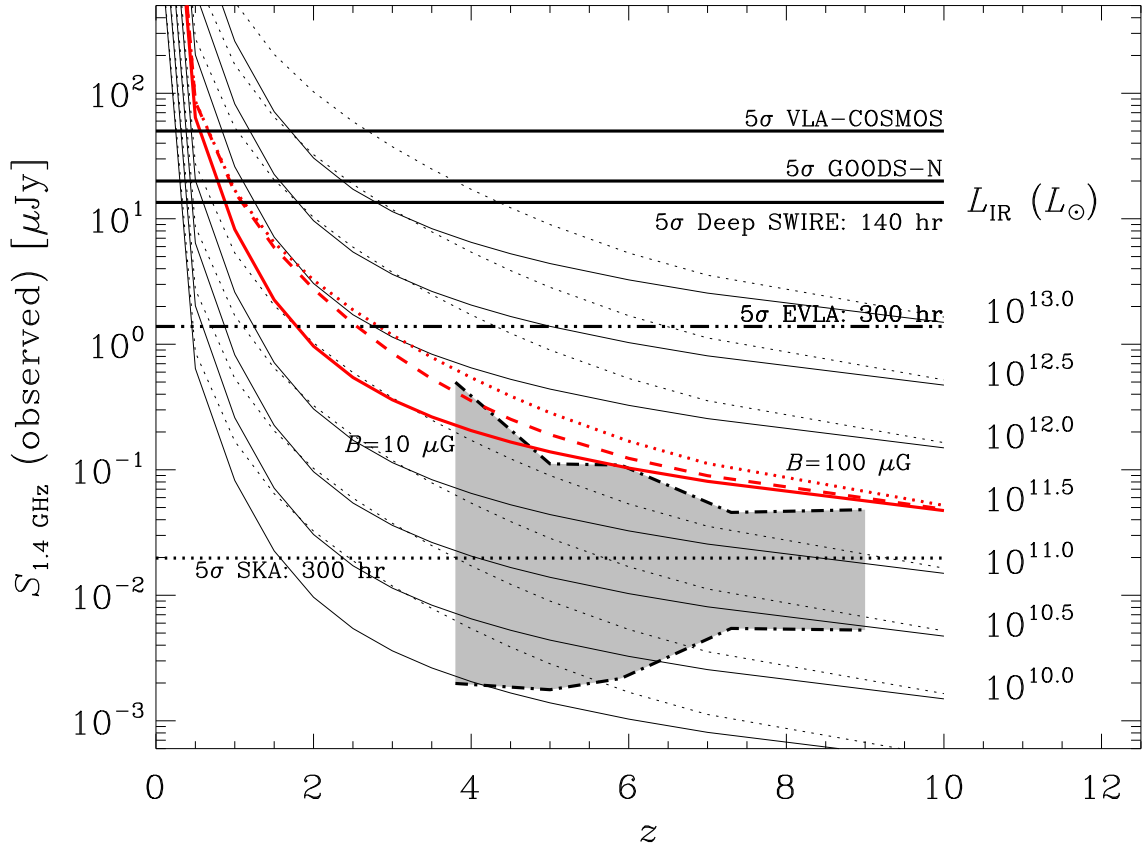}
\caption{
The expected observed-frame 1.4~GHz flux density for galaxies of varying IR luminosities assuming the FIR-radio correlation (i.e. $q_{\rm IR}=2.64$) as a function of redshift for nominal galaxy parameters (Table \ref{tbl-1}).    
Additional energy losses to CR electrons arising from IC scattering off of the CMB, whose radiation energy density scales as $\sim(1+z)^{4}$, are included.   
Flux densities, corrected for bandwidth compression, are plotted assuming internal magnetic field strengths of 10~$\mu$G ({\it solid}-line) and 100~$\mu$G ({\it dotted}-line); 
at high~$z$ the non-thermal emission from galaxies is highly suppressed even if $B\sim100~\mu$G, leaving only the thermal component detectable.   
The case of a moderate LIRG ($L_{\rm IR} \approx 3\times10^{11}~L_{\sun}$, implying a star formation rate of $\sim 50~M_{\sun}~{\rm yr}^{-1}$) is highlighted for which a 50~$\mu$G magnetic field strength ({\it dashed}-line) is also shown.    
The sensitivities of existing deep radio continuum surveys using the VLA, the expected depth of the EVLA, and suggested depth for the SKA (Table \ref{tbl-2}), which would detect typical LIRGs (i.e. galaxies forming stars at rates $\ga 25~M_{\sun}~{\rm yr}^{-1}$) at all $z$ if present, are shown.  
At this depth the SKA will be sensitive to galaxies lying in the shaded region which indicates the expected flux density range for galaxies included in the $z\sim$ 4, 5, 6, 7, and 9 UV luminosity function studies of \citet{rb07,rb08}.  
\label{fig-3}}
\end{figure*}

For a typical star-forming galaxy, CR electrons primarily lose their energy due to synchrotron and IC processes \citep[e.g.][]{jc92}, although escape also plays a role \citep[e.g.][]{hb93}.            
Let us assume that CR electrons propagating with a pitch angle $\phi$ in a magnetic field of strength $B$ have isotropically distributed velocities such that \(<\sin^{2}\phi> = \twothirds\) leading to \(B_{\perp} \approx 0.82~B\).
A CR electron having energy $E$ will emit most of its energy at a critical frequency $\nu_{\rm c}$ where  
\begin{equation} 
\label{eq-nuBE}
 \left(\frac{\nu_{\rm c}}{\rm GHz}\right) = 1.3\times10^{-2}
  \left(\frac{B}{\rm \mu G}\right)
  \left(\frac{E}{\rm GeV}\right)^{2}.
\end{equation}

The energy loss of CR electrons by synchrotron radiation goes as $dE/dt \propto U_{B}E^{2}$, where \(U_{B} = B^{2}/(8\pi)\) is the magnetic field energy density of the galaxy.  
Using Equation \ref{eq-nuBE}, the synchrotron cooling timescale, \(\tau_{\rm syn} \equiv E/|dE/dt|_{\rm syn}\), for CR electrons can be expressed as   
\begin{equation}
\label{eq-tsync_ub}
  \left(\frac{\tau_{\rm syn}}{\rm yr}\right) \sim 5.7\times10^{7}
  \left(\frac{\nu_{\rm c}}{\rm GHz}\right) ^{-1/2}
  \left(\frac{B}{\rm \mu G}\right)^{1/2}
  \left(\frac{U_{B}}{10^{-12}~{\rm erg~cm^{-3}}}\right)^{-1}.
\end{equation}
Naturally, the synchrotron cooling timescale for CR electrons can be rewritten as 
\begin{equation}
  \label{eq-tsync}
  \left(\frac{\tau_{\rm syn}}{\rm yr}\right) \sim 1.4\times10^{9}
  \left(\frac{\nu_{\rm c}}{\rm GHz}\right) ^{-1/2}
  \left(\frac{B}{\rm \mu G}\right)^{-3/2}.
\end{equation}
Similarly, the energy loss of CR electrons due to IC scattering goes as $dE/dt \propto U_{\rm rad}E^{2}$ where $U_{\rm rad}$ is the radiation field energy density of the galaxy.  
Equation \ref{eq-nuBE} can therefore be used again to write the IC cooling timescale, \(\tau_{\rm IC} \equiv E/|dE/dt|_{\rm IC}\), as
\begin{equation}
\label{eq-tIC}
  \left(\frac{\tau_{\rm IC}}{\rm yr}\right) \sim 5.7\times10^{7}
  \left(\frac{\nu_{\rm c}}{\rm GHz}\right) ^{-1/2}
  \left(\frac{B}{\rm \mu G}\right)^{1/2}
  \left(\frac{U_{\rm rad}}{10^{-12}~{\rm erg~cm^{-3}}}\right)^{-1}. 
\end{equation}
For GeV electrons considered here,  
the bulk of losses arise from interactions with IR/optical photons, which dominate $U_{\rm rad}$.

The effective cooling timescale for CR electrons due to synchrotron and IC losses is 
\begin{equation}
\label{eq-cool_shrt}
\tau_{\rm cool}^{-1} = \tau_{\rm syn}^{-1} + \tau_{\rm IC}^{-1},
\end{equation}
which, by combining Equations \ref{eq-tsync_ub} and \ref{eq-tIC},  we
can express as
\begin{equation}
\label{eq-cool}
  \left(\frac{\tau_{\rm cool}}{\rm yr}\right) \sim 5.7\times10^{7}
  \left(\frac{\nu_{\rm c}}{\rm GHz}\right) ^{-1/2}
  \left(\frac{B}{\rm \mu G}\right)^{1/2}
  \left(\frac{U_{B}+U_{\rm rad}}{10^{-12}~{\rm
  erg~cm^{-3}}}\right)^{-1}.
\end{equation}

If CR electrons do not lose all of their energy as they propagate through the ISM of a galaxy, they may eventually escape the system.  
In simple diffusion models, the propagation of CR electrons is usually characterized by an empirical, energy-dependent diffusion coefficient, $D_{E}$ \citep[e.g.][]{ginz80}.
Values of $D_{E}$ has been found to be around \(4-6\times 10^{28}~{\rm cm^{2}~s^{-1}}\) for $\la$GeV CRs by fitting diffusion models with direct measurements of CR nuclei  (i.e. secondary-to-primary ratios like Boron-to-Carbon) within the Solar Neighborhood \citep[e.g.][]{fj01,im02,dm02}.
While this empirically measured value is for CR nuclei within the Milky Way, it has been found to be consistent with inferred diffusion coefficients for CR electrons both radially along field lines (\(\sim10^{29}~{\rm cm^{2}~s^{-1}}\)) and vertically across field lines (\(\sim10^{28}~{\rm cm^{2}~s^{-1}}\)) in the thin disks of galaxies \citep[i.e. NGC~891 and NGC~4631;][]{dlg95}.  
\citet{vh09} also report a diffusion coefficient of \(\sim2\times10^{29}~{\rm cm^{2}~s^{-1}}\) in the halo of NGC~253.  
Similar values near \(\sim10^{29}~{\rm cm^{2}~s^{-1}}\) are also found in hydrodynamic simulations of bubble/super-bubble induced galaxy outflows \citep[e.g.][]{pr05,er07}.
Using the average value of the diffusion coefficient obtained through direct measurements of CR nuclei, which is quite similar to that inferred for radial and vertical diffusion coefficients for CR electrons, the diffusion coefficient is expressed as   
\begin{equation}
\label{eq-DE}
\left(\frac{D_{E}}{\rm cm^{2}~s^{-1}}\right) \sim \Bigg\{
\begin{array}{cc}
5 \times 10^{28}, & E < 1~{\rm GeV}\\
5 \times 10^{28}(\frac{E}{{\rm GeV}})^{1/2}, & E \geq 1~{\rm GeV}.
\end{array}
\end{equation}
Then, for for the case of random walk diffusion, such that \(\tau_{\rm esc} = l_{\rm esc}^{2}/D_{\rm E}\),  where $l_{\rm esc}$ is the escape scale-length, CR electrons will escape a galaxy in   
\begin{eqnarray}
\left(\frac{\tau_{\rm esc}}{\rm yr}\right) \sim \Bigg\{
\begin{array}{cc}
6.0\times10^{6}\left(\frac{l_{\rm esc}}{\rm kpc}\right)^{2}, & E < 1~{\rm GeV}\\
2.0\times10^{6}
	\left(\frac{l_{\rm esc}}{\rm kpc}\right)^{2}
	\left(\frac{\nu}{\rm GHz}\right)^{-1/4}
	\left(\frac{B}{\rm \mu G}\right)^{1/4}, & E \geq1~{\rm GeV}, 
\end{array}
\end{eqnarray}
where Equation \ref{eq-nuBE} has been used to express the electron energy dependence in terms of its synchrotron emitting frequency for a given magnetic field strength.  

Most edge-on galaxies possess thick synchrotron disks with scale-heights ranging between $1-3$~kpc \citep{ul04}.  
In the case of equipartition between the energy density of cosmic rays and the magnetic field, the CR scale-height will be $(3+\alpha)/2\approx 2$ times that of the synchrotron scale-height for a synchrotron spectral index of $\alpha \approx 0.8$ \citep[e.g.][]{nkw97}.      
Then, assuming a magnetic field strength of 10~$\mu$G and a CR scale-height of $\sim$3~kpc (synchrotron scale-height of 1.5~kpc), the typical escape time of a CR electron emitting at 1.4~GHz is $\sim$$3\times10^{7}$~yr.  
This is comparable to the expected cooling time of such electrons to synchrotron and IC processes, assuming that $U_{\rm rad} = 10^{-12}$~erg~cm$^{-3}$, indicating that normal galaxies do {\it not} behave like calorimeters.  

Vertical diffusion of CRs, eventually leading to their escape into intergalactic space due to either open magnetic field lines or Parker instabilities \citep{ep66}, was looked at in detail by \citet{hb93}.  
Their proposed physical model for the FIR-radio correlation includes a leaky-box  confinement scheme in which the escape scale-length of CR electrons is nearly independent of ISM density and magnetic field strength, but scales with disk scale-height to keep a constant IR/radio ratio.  
While escape may play a role, injecting scatter into the observed IR/radio ratios in larger spirals, it is noteworthy that for the case of irregular galaxies, which generally lack dense ISM and magnetic field to keep CR electrons trapped once leaving their initial clouds around SNRs, escape appears to play a much more critical role pushing such galaxies off of the correlation \citep[e.g.][]{jc05,jc06,ejm08}.

\subsection{Essential Physics: Starbursting Galaxies}
Additional energy-loss terms that may become increasingly important in the case of  galaxies hosting strong starbursts are now considered.  
In such systems, whose energetics and ISM may be vastly different, ionization, bremsstrahlung, and adiabatic cooling through advection out of a galaxy by galactic-scale winds, can all play a significant role in cooling CR electrons.  
Ionization losses begin to dominate over synchrotron and IC losses for CR electron energies below about $\sim$1.3~GeV such that,  
\begin{eqnarray}
  \begin{array}{ll}
 \left(\frac{\tau_{\rm ion}}{\rm yr}\right) &
 \sim 4.1\times10^{9}
 \left(\frac{E}{\rm GeV}\right)
 \left(\frac{n_{\rm ISM}}{\rm cm^{-3}}\right)^{-1}\\  
& \left[3\ln\left(\frac{E}{\rm GeV}\right)+42.5\right]^{-1}.
\end{array}
\end{eqnarray}
Combining this with Equation \ref{eq-nuBE} yields, 
  \begin{eqnarray}
\begin{array}{ll}
 \left(\frac{\tau_{\rm ion}}{\rm yr}\right) & 
 \sim 3.6\times10^{10}
 \left(\frac{\nu}{\rm GHz}\right)^{1/2}
 \left(\frac{B}{\rm \mu G}\right)^{-1/2}
 \left(\frac{n_{\rm ISM}}{\rm cm^{-3}}\right)^{-1}\\
&  \left(\frac{3}{2} \left[\ln\left(\frac{\nu}{\rm GHz}\right) - \ln\left(\frac{B}{\rm \mu G}\right)\right]+49\right)^{-1}.
\end{array}
\end{eqnarray}
Thus, if magnetic fields are sufficiently large in starbursting systems (i.e. $\ga$1~mG), as suggested by \citet{tt06}, electrons radiating GHz synchrotron continuum emission will have energies $\la $0.3~GeV and ionization losses should dominate over synchrotron cooling likely leading to depressed synchrotron emission (see Figure 1).  

Furthermore, bremsstrahlung losses are probably negligible in normal galaxies given that the CR electron mean free path is $\sim$5~g~cm$^{-2}$ \citep{gm77}, an order of magnitude smaller than the $\sim$60~g~cm$^{-2}$ for the radiation length of the ISM.  
However, for a sufficiently dense ISM, bremsstrahlung losses may exceed ionization losses for electrons at energies above $\sim$1.1~GeV such that, for neutral hydrogen, 
 \begin{equation}
 \left(\frac{\tau_{\rm brem}}{\rm yr}\right) \sim 8.6\times 10^{7}
 \left(\frac{n_{\rm ISM}}{\rm cm^{-3}}\right)^{-1}.  
 \end{equation}
Additionally, if these starbursts are compact, the adiabatic expansion losses to CR electrons may become more rapid due to the advection out of each system with a galactic wind, having speed $v_{w}$, where the adiabatic lifetime goes as, 
\begin{equation}
\left(\frac{\tau_{\rm ad}}{\rm yr}\right) \sim 1.0\times10^{9}
	\left(\frac{h}{\rm kpc}\right)
	\left(\frac{v_{w}}{\rm km~s^{-1}}\right)^{-1}.
\end{equation}
In either case, if the physical conditions of galaxies at higher redshifts are typically well described by compact starbursts, the above mentioned energy-loss processes will become important relative to synchrotron losses for GHz radiating CR electrons.  

To demonstrate this I plot the cooling timescales of CR electrons associated with each of the above mentioned energy-loss processes versus electron energy in Figure \ref{fig-1}.  
In the left panel of Figure \ref{fig-1}  
parameters typical for normal star-forming galaxies are assumed: $B=10~\mu$G, $U_{\rm rad} = 10^{-12}$~erg~cm$^{-3}$, $n_{\rm ISM} = 1$~cm$^{-3}$, and $l_{\rm esc} = 3$~kpc. 
Losses due to advection by a galactic wind are ignored.  
At GHz frequencies synchrotron and IC losses are clearly the dominant energy-loss terms along with escape.  
In the right panel of Figure \ref{fig-1} the same 
cooling timescales are plotted as in the left panel, however the case of a compact starburst having a 1~mG magnetic field is considered.  
Assuming a flux freezing scaling of $B\propto \sqrt{n_{\rm ISM}}$ \citep[e.g.][]{ruz88, hb93, nb97, cr99}, it follows that $n_{\rm ISM} = 10^{4}$~cm$^{-3}$.  
The same ratio of $U_{B}/U_{\rm rad}$ as was used in the left panel is also assumed, setting $U_{\rm rad} = 10^{-8}$~erg~cm$^{-3}$.  
Energy losses via advection by a starburst-driven galactic wind, having a velocity of $v_{w}=300~$km~s$^{-1}$ \citep[i.e. typical of $z\sim$3 Lyman break galaxies;][]{as03} and assuming a disk scale-height $h= 1$~kpc, are also shown.  
At GHz frequencies it is found that synchrotron radiation and IC scattering are no longer the dominant energy-loss mechanisms.  
Ionization and bremsstrahlung losses now dominate the cooling of CR electrons contributing to the GHz emission from galaxies.  
While such strong fields are present in starbursts on small scales, as Zeeman splitting measurements of OH megamasers in ULIRGs show line-of-sight field strengths ranging from $\approx0.5-18$~mG in individual masing components \citep{tr08}, 
given that local starburst galaxies follow the FIR-radio correlation, this may suggest that their average magnetic fields never reach $\sim$mG strengths since the depressed synchrotron emission at GHz frequencies should yield larger than nominal IR/radio ratios, which is not observed.  

\subsection{Radio Spectra of Galaxies}  
Let us now see how the combination of these energy-loss processes translates into the radio spectra of galaxies.  
For an electron injection spectrum $Q(E) = \kappa E^{-p}$, the number density of particles per unit energy is expressed as \citep[e.g][]{ml94}
\begin{equation}
N(E) = \frac{\kappa E^{-(p-1)}}{(p-1)b(E)}
\end{equation}
where $b(E) = -dE/dt$ is the total energy losses for electrons including escape.  
The injected electrons, if shock-accelerated by SNRs,  will have a spectrum typically characterized by $p_{\rm inj} \approx 2.4$.   
However, as the CR electrons interact with the ISM, diffusive losses will modify the spectrum to have an index of $p\approx$2.8 \citep[e.g.][]{bs93,jb09}, which is assumed for the model.
Then, the non-thermal radio continuum emission (i.e. the energy losses to CR electrons by synchrotron radiation relative to the total energy losses) is given by 
\begin{equation}
S_{\nu}^{\rm NT} \propto E/\tau_{\rm sync} N(E), 
\end{equation}
where emission at frequency $\nu$ depends on CR electron energy $E$ according to Equation \ref{eq-nuBE}.  
For a constant electron temperature, the amount of thermal radio continuum emission goes as 
\begin{equation}
S_{\nu}^{\rm T} \propto \nu^{-0.1}.  
\end{equation}
Assuming $B \approx 10~\mu G$, $U_{B} \approx U_{\rm rad}$, and $n_{\rm ISM} \approx 1~{\rm cm}^{-3}$, the expected radio continuum spectrum is plotted in Figure \ref{fig-2} along with an IR dust emission spectrum scaled such that its infrared (IR; $8-1000~\mu$m)  flux is $10^{-13}$~W~m$^{-2}$.  
The dust emission spectrum is taken from the SED libraries of \citet{dh02} using the template which  best fit typical star-forming galaxies in the {\it Spitzer} Infrared Nearby Galaxies Survey \citep[SINGS;][]{rk03}.  
The thermal radio spectrum was then normalized by equating the relations between the ionizing photon rate and the thermal radio continuum \citep{jc92} and IR luminosity  \citep{rk98} resulting in 
\begin{eqnarray}
\label{eq-stir}
\begin{array}{ll}
\left(\frac{S_{\nu}^{\rm T}}{\rm W~m^{-2}~Hz^{-1}}\right) &
	\sim 6.6\times10^{-17} 
	\left(\frac{T_{e}}{\rm 10^{4}~K}\right)^{0.45}\\ 
&	\left(\frac{\nu}{\rm GHz}\right)^{-0.1}
	\left(\frac{F_{\rm IR}}{\rm W~m^{2}}\right)
\end{array}
\end{eqnarray}
where $T_{e} = 10^{4}$~K is the assumed electron temperature for an H{\sc ii} region.  

Next, the non-thermal component was scaled such that the thermal radio fraction is $\sim$10\% at 1.4~GHz, which is typically found for star-forming systems \citep{cy90,nkw97}.    
The combination of these assumptions naturally leads to the nominal FIR-radio correlation where  
\begin{equation}
\label{eq-q}
q \equiv \log~\left(\frac{F_{\rm FIR}}{3.75\times10^{12} ~{\rm W~m^{-2}}}\right) - 
	\log~\left(\frac{S_{1.4~{\rm GHz}}}{\rm W ~m^{-2} Hz^{-1}}\right)
\end{equation}
is $\approx$$2.34\pm 0.26$~dex \citep{yrc01}.
Replacing the FIR flux with the total IR flux in Equation \ref{eq-q} leads to an average $q_{\rm IR} \approx 2.64 \pm0.26$~dex \citep{efb03}.  
For these parameters, it is also found that a least squares fit to the non-thermal radio spectrum yields a spectral index of $\alpha_{\rm NT}\sim 0.8$, similar to what is observed in normal star-forming galaxies \citep[e.g.][]{nkw97}.

To keep a fixed ratio between the FIR and radio emission of galaxies requires that the ratio of synchrotron to total energy losses for CR electrons to be nearly constant among each system.  
Or, in the case of normal star-forming galaxies, where CR electron energy losses are dominated by synchrotron losses and IC scattering off a galaxy's interstellar radiation field (ISRF), the FIR-radio correlation is preserved under the condition where
\begin{equation}
\label{eq-ucmb}
\frac{U_{B}}{U_{\rm rad} + U_{\rm CMB}} \ga 1, 
\end{equation}
where $U_{\rm CMB}$ is the radiation field energy density of the cosmic microwave background (CMB).  
In the local Universe, IC losses off of the CMB are negligible to those occurring off the ISRF within a galaxy, however this is not necessarily the case at higher redshifts.

\section{Results: Expectations for Deep Radio Continuum Surveys}
I now explore implications of the above physical arguments for evolution in the observed radio continuum emission from galaxies at increasing redshifts.  
I also report on how this relates to an expected evolution in the FIR-radio correlation; 
at present, it is assumed that any change in the FIR-radio correlation arises solely from changes to the radio continuum emission of galaxies.   
Variations arising from differences in the IR emission of galaxies are discussed below in $\S$\ref{sec-dust}.  
Implications for how various galaxy properties, including their star formation activity and magnetic field properties, can be inferred from the discrepancies between expected and observed IR/radio ratios are discussed in $\S$4.

\begin{figure}
\plotone{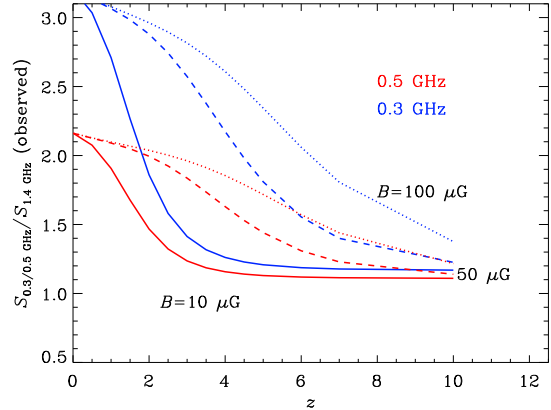}
\caption{
The ratio of observed-frame 0.3 and 0.5~GHz  to 1.4~GHz flux densities plotted as a function of redshift in blue and red, respectively.  
These estimates include the expected suppression of non-thermal radio continuum as IC scattering of the CMB become increasingly important at higher redshifts (see $\S$\ref{sec-ntsup} for details about the calculations.)  
Different cases for the intrinsic magnetic field strength of galaxies are considered: 10~$\mu$G ($solid$ lines), 50~$\mu$G ($dashed$ lines), and 100~$\mu$G ($dottted$ lines).  
Unless magnetic field strengths are very strong ($\ga10~\mu$G), only marginal sensitivity gains exist by making lower frequency radio observations to detect galaxies at $z \ga 2$.   
\label{fig-4}}
\end{figure}

\subsection{Suppression of Non-Thermal Emission by the CMB \label{sec-ntsup}}
As already shown, the FIR-radio correlation relies on a fixed ratio between synchrotron and the total energy losses of CR electrons.    
At $z =0$, $U_{\rm CMB} \sim 4.2\times10^{-13}~{\rm erg~cm^{-3}}$ which is significantly smaller than the radiation field energy density of the Milky Way (i.e. $U_{\rm MW} \sim 10^{-12} ~{\rm erg~cm^{-3}}$).  
Thus, CR electron energy losses from IC scattering off the CMB are negligible at low redshifts.  
However, $U_{\rm CMB} \propto (1+z)^{4}$ making such losses increasingly important with redshift.  
For instance, by $z\sim3$, $U_{\rm CMB} \sim 1.1\times 10^{-10}~{\rm erg~cm^{-3}}$; equating this to $U_{B}=B^{2}/(8\pi)$ results in a corresponding magnetic field strength of $\sim$50~$\mu$G, nearly an order of magnitude larger than the ambient field strength in the Solar Neighborhood.  
Consequently, the non-thermal component of a galaxy's radio continuum emission will be increasingly suppressed with increasing redshift, {eventually resulting in only the thermal component being detectable.}  

Assuming an intrinsic FIR-radio correlation for galaxies with $q_{\rm IR} \approx 2.64$ \citep{efb03}, the expected observed-frame 1.4~GHz flux density for star-forming galaxies having a range of IR ($8-1000~\mu$m) luminosities as a function of redshift are estimated (Figure \ref{fig-3}).     
These estimates rely on the results of the previous section where, on a theoretical basis, a realistic radio spectrum was constructed.    
It is assumed that the radio continuum emission is comprised of two components, thermal (free-free) and non-thermal (synchrotron) emission, both of which can be expressed as power-laws such that $S_{\nu} \propto \nu^{-\alpha}$, where $\alpha_{\rm T} \sim 0.1$ for the thermal (free-free) component and $\alpha_{\rm NT} \sim 0.8$ for the non-thermal component (Table \ref{tbl-1}).  
A thermal fraction of $\sim$10\% at 1.4~GHz is also assumed.  
And a $(1+z)^{-1}$ correction factor to account for bandwidth compression as a function of redshift is included.
Increased energy losses to CR electrons from IC scattering as $U_{\rm CMB}$ increases with redshift are explicitly included.  
Figure \ref{fig-3} illustrates two cases per IR luminosity in which the galaxy has a 10 and 100~$\mu$G field, indicated by the solid and dotted lines, respectively.     
The case of a moderately bright LIRG ($L_{\rm IR} \approx 3\times10^{11}~L_{\sun}$) is highlighted in red,  and the case of a 50~$\mu$G field strength shown using a dashed line.  

{For the case of a moderately bright LIRG, it is clearly shown that its detection in the radio continuum may significantly rely on the strength of its magnetic field between $1 \la z \la 6$ due to the suppression of its synchrotron emission by increased IC losses off of the CMB.  
For example, a galaxy having $B = 10~\mu$G at $z\sim3$ will also be detected at $z\sim4.5$ if its magnetic field were 100~$\mu$G.
At higher redshifts it appears that the strength of the magnetic field plays less of a role since IC losses off of the CMB begin to completely suppress a galaxy's non-thermal emission, making only its thermal radio emission detectable.  
It is also shown that a moderately bright LIRG at $z\sim10$ should have a 1.4~GHz flux density of $\sim$40~nJy almost entirely arising from free-free emission.  
Using the star formation rate conversion of \citet{rk98}, this implies that all galaxies forming stars at a rate of  $\ga50~M_{\sun}~{\rm yr}^{-1}$ should be detectable out to $z\sim10$ if such a population of galaxies exists at these early epochs and this sensitivity is achievable.  

Due to the relatively IR/radio negative spectral slope of a galaxy's non-thermal radio continuum emission, the case of observing at lower frequencies is considered to see what sensitivity gains might exist for detecting galaxies at high-$z$.  
The expected observed-frame flux densities at 0.3 and 0.5~GHz are estimated.     
The ratio between the observed-frame 0.3 and 0.5 to 1.4~GHz flux densities as a function of redshift is plotted in Figure \ref{fig-4}.  
The cases for different intrinsic $B$-field strengths among galaxies (i.e. 10, 50, and 100~$\mu$G) are again considered.  
For galaxies having magnetic field strengths of $\sim$10~$\mu$G it appears that there is little advantage for surveys at lower frequencies since, by $z\sim2.5$, sensitivity gains are less than $\sim$50\%.  
If a galaxy's magnetic field is quite strong, however, being $\sim$100~$\mu$G, observations at 0.3~GHz should be more sensitive to normal star-forming galaxies than 1.4~GHz observations by a factor of $\ga$2 out to $z\sim6$.  
While the possibility of increased magnetic field strengths in galaxies at higher redshifts is considered, there is not a strong physical basis to expect this as the leading theory of magnetic field generation, the mean-field dynamo model, predicts large-scale regular magnetic fields to be weaker in the past rather than stronger  \citep[e.g.][]{rb96}.

\subsection{Expected Evolution in the FIR-Radio Correlation with Redshift from CMB Effects}
As already shown, the non-thermal component of a galaxy's radio continuum emission is increasingly suppressed as a function of redshift due to the increasing importance of IC scattering off of the CMB.  
This effect should be directly reflected in changes to a galaxy's IR/radio ratio.  
Assuming that the intrinsic IR/radio ratio of star-forming galaxies is not unique to $z\sim0$ systems, the FIR-radio correlation should evolve in a very predictable way as a function of redshift arising from the increase in the energy density of the CMB.  
This is illustrated in Figure \ref{fig-5}.   
When the non-thermal emission is completely suppressed by IC scattering off of the CMB, observed IR/radio ratios are expected to simply approach the ratio between the FIR and thermal radio continuum.    

As already stated, different CR electron cooling processes may become increasingly more important than synchrotron radiation in galaxies hosting strong starbursts.  
For such cases, the non-thermal component may diminish even more quickly than in the calculation provided here, resulting in observed IR/radio ratios for which only the thermal radio continuum emission is detected (i.e. $q_{\rm IR}$ ratios which are a factor of $\sim$10 higher than the nominal value).  
However, it is also possible that other physical processes could work in the opposite direction to increase the synchrotron emissivity, such as an increase to the acceleration efficiency of CR electrons within SNRs, as well as additional synchrotron emission arising from an increase to secondary positrons and electrons.  
These possibilities, though speculative since such local starbursts do not seem to exhibit such variations in their IR/radio ratios, are discussed in $\S$\ref{sec-acc} and \ref{sec-sec}, respectively.

\begin{figure}
\plotone{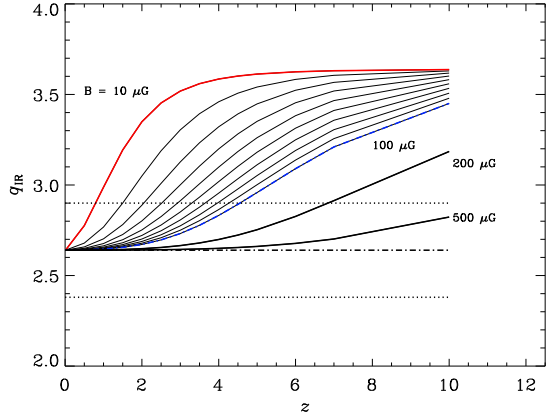}
\caption{
The expected rest-frame IR/radio ratios for a galaxy having $q_{\rm IR}=2.64$ at $z=0$ as IC losses off of the CMB become increasingly important as a function of redshift.  
Each track corresponds to a different internal magnetic field strength for the galaxy.  
As $U_{\rm CMB} \gg U_{B}$, the IR/radio ratio approaches the limit where only thermal (free-free) radiation contributes to the observed radio continuum emission.  
The average local $q_{\rm IR}$ values (2.64~dex; {\it dot-dashed} line) and the $\pm 1~\sigma$ scatter ({\it dotted}-line) are shown.  
\label{fig-5}}
\end{figure}

\subsection{Scientific Goals and Technical Specifications for Deep Radio Continuum Surveys using the SKA}  
Detecting moderately bright star-forming galaxies at redshifts beyond $z\sim3$ requires extremely sensitive continuum imaging only expected with the advent of the SKA. 
Figure \ref{fig-3} shows the achieved point-source sensitivities for a number of existing 1.4~GHz radio continuum surveys including: 
VLA-COSMOS \citep[][$\sigma_{\rm RMS}\approx 50~\mu$Jy]{eva07}, 
GOODS-N (G. Morrison 2009, in preparation, 5$\sigma_{\rm RMS}\approx 20~\mu$Jy, and 
the Deep SWIRE Field \citep[][ 5$\sigma_{\rm RMS}\approx $13.5~$\mu$Jy]{om08}.  
The expected point-source sensitivity of the EVLA after 300~hr ($\sim$1 Ms) is also plotted (5$\sigma_{\rm RMS}\approx 1.4~\mu$Jy).  
As shown, even the EVLA will not be able to detect individual M~82-like objects, having IR luminosities $\sim$$4\times10^{10}~L_{\sun}$, at $z\ga1$.   

As discussed in $\S$\ref{sec-ntsup}, a 5$\sigma$ point-source sensitivity of $\approx$40~nJy is required for the detection of a moderately bright LIRG (i.e. $L_{\rm IR} \approx 3\times10^{11}~L_{\sun}$), forming stars at $\sim$50~$M_{\sun}~{\rm yr}^{-1}$ at $z\sim10$.  
However, it is unclear as to whether there will be a significant population of dusty star-forming galaxies at  these redshifts.  
At present, only rest-frame UV studies relying on the "dropout" technique are able to probe normal galaxies at such high redshifts.  
In Figure \ref{fig-3} the estimated observed-frame 1.4~GHz flux densities, based on the results from the UV ($\approx$1600~\AA) luminosity function work of \citet{rb07,rb08}, are shown.  
These authors use space- and ground-based data over both {\it Hubble} ultra-deep fields and GOODS fields to search for star-forming galaxies at $z\sim$4, 5, 6, 7, and 9.  
The UV luminosities, over the full range per redshift bin as given in Table 5 of \citet{rb07} and Table 3 of \citet{rb08}, were corrected for dust following \citet{rb07} such that $A_{1600} =$ 1.1~mag, 0.6~mag, and 0.4~mag for $z\sim4$, $z\sim5$, and $z\ga6$ galaxies, respectively.  
Corresponding observed-frame radio flux densities were then estimated by first converting the extinction-corrected UV luminosities to IR luminosities following \citet{rk98}, and then using the FIR-radio correlation assuming a galaxy magnetic field strength of 10~$\mu$G.  

These considerations suggest that the criterion of a point-source sensitivity of $\approx$40~nJy  
will only skim the brightest sources among this population of galaxies.  
Beyond $z\ga7$, detection of these galaxies will be marginal at best.  
However, it is possible that the assumed dust corrections significantly underestimate the true star formation rates of these sources as they are based on an empirical relation between the UV continuum  slope and extinction in local starbursting galaxies \citep[e.g.][]{gm99}.  
While this prescription works well for $z\sim2$ LIRGs older than $\sim$100~Myr \citep{nr06}, it has also been shown to fail for many other high-$z$ systems \citep[e.g.][]{jg02,nr06,ejm09a}.  
Furthermore, the $4 \la z \la 9$ UV luminosity function work suffers dramatically from the effects of cosmic variance.  
In any case, it appears quite advantageous for the SKA sensitivity requirement be a factor of $\sim$2 better, namely a 5$\sigma$ point-source sensitivity of $\approx 20$~nJy (i.e. $\sigma_{\rm RMS} \approx 4$~nJy).

The sensitivity for the SKA is approximated using the radiometer equation such that \(\sigma_{\rm RMS} \approx 2k_{B}/\sqrt{2~BW~t_{\rm int}}~(A_{\rm eff}/T_{\rm sys})^{-1}\) where $k_{B}$ is the Boltzmann constant, BW is the bandwidth, $t_{\rm int}$ is the integration time, $A_{\rm eff}$ is the effective collecting area of the array, and $T_{\rm sys}$ is the system temperature.    
Assuming a bandwidth of  $\sim$1~GHz (comparable to what existing telescopes are delivering) and a reasonably long integration (i.e. 300~hr), I find a requirement for the SKA of  $A_{\rm eff}/T_{\rm sys} \sim 15000~{\rm m^{2}~K^{-1}}$ to reach this sensitivity such that, 
\begin{equation}
\left(\frac{\sigma_{\rm RMS}}{\rm nJy}\right) \approx 4~
\left(\frac{\rm BW}{\rm GHz}\right)^{-1/2}
\left(\frac{t_{\rm int}}{\rm 300~hr}\right)^{-1/2}
\left(\frac{A_{\rm eff}/T_{\rm sys}}{\rm 15000~{\rm m^{2}~K^{-1}}}\right)^{-1}.  
\end{equation}
With these specifications, the SKA will be sensitive to all galaxies forming stars at a rate of $\ga25~M_{\sun}~{\rm yr}^{-1}$.  
The $5\sigma_{\rm RMS} \approx 20$~nJy is also plotted in Figure \ref{fig-3}.  

\subsubsection{
Technical Challenges for Deep SKA Continuum Fields}
In order to achieve such highly sensitive radio imagery, two major technical challenges must be considered in the SKA design to ensure that systematic noise does not limit the usability of such imaging;  natural confusion from overlapping sources in the field and insufficient dynamic range to identify sources properly.  
Both effects were recently considered by \citet{jc09}, who found that they should not limit deep continuum surveys using the SKA at nJy depths.  

While the angular size distribution of the sub-$\mu$Jy population remains uncertain, most faint radio sources are likely coextensive with faint star-forming galaxies at $z\sim1$.  
These sources are therefore probably smaller than $\sim$1\arcsec setting the natural confusion limit to be the flux density corresponding to $\sim$30~arcsec$^{-2}$ per source.  
Accordingly, the synthesized beam of the SKA should be $\theta_{\rm S} \la 1 \arcsec$ to keep instrumental confusion below the natural confusion limit, requiring baselines up to $\sim$50~km at 1.4~GHz \citep{jc09}.  
\citet{rw08} also estimate the expected 1.4~GHz source sizes at $10-100$~nJy by taking the galaxy disk size distributions measured from deep HST imaging, where the object density is in excess of $10^{6}$ objects~deg$^{-2}$.  
Their findings suggest that the synthesized beam of the SKA be even sharper, reaching $\theta_{\rm S} \la $0\farcs3, to avoid instrumental confusion and  requiring baselines up to $\sim$150~km at 1.4~GHz.  
The 1.4~GHz natural confusion is estimated to be $\sigma_{\rm c} \la 3$~nJy, though with large uncertainty \citet{jc09}, suggesting that from the point-source sensitivity calculation above, the SKA should not be limited by natural confusion.  

\begin{deluxetable}{lc}
\tablecaption{Assumptions for Expected Radio Flux Densities \label{tbl-1}}
\tablewidth{0pt}
\tablehead{
\colhead{Parameter} & \colhead{Value}
}
\startdata
$q_{\rm IR}$ & 2.64 \\
$\alpha_{T}~^{\dagger}$ & 0.1 \\
$\alpha_{NT}~^{\dagger}$ & 0.8\\
$f_{\rm T}^{\rm 1.4~GHz}$ & 0.10\\
$U_{\rm CMB} (z = 0)$ & $4.2 \times 10^{-13}{\rm ergs~cm^{-3}}$
\enddata
\tablecomments{$^{\dagger}$: The thermal and non-thermal radio continuum is assumed to follow power-laws of the form $S_{\nu}\propto \nu^{-\alpha}$, where $\alpha_{\rm T}$ and $\alpha_{\rm NT}$ are the assumed thermal and non-thermal spectral indices.}
\end{deluxetable}

Coupled with the requirement of exquisite sensitivity is achieving an unprecedented dynamic range to actually  make use of such deep imagery.  
Typically, the dynamic range is defined as $D \equiv S_{\rm max}/\sigma_{\rm RMS}$, the ratio of the strongest peak ($S_{\rm max}$) to RMS fluctuations in regions not containing sources, and describes an image dominated by an extremely strong point source.  
At present, the highest dynamic range achieved has been $\sim$60~dB.  
For example, a dynamic range of $\sim$60~dB was obtained with Westerbork Synthesis Radio Telescope (WSRT) observations over a 10\arcmin~ region containing a single bright source (i.e. 3C~147).  
At the desired nJy sensitivity, the large primary beam of the SKA will contain many sources having strengths larger than that of the radiometer noise.  
Thus, the dynamic range required by the SKA can be estimated using the typical flux density for the strongest source in a field, which for a primary beam $\Omega_{\rm b} \approx 1$~deg$^{2}$, is $\sim$0.16~Jy at 1.4~GHz \citep{jc09}.  
Using the required RMS noise of $\sigma_{\rm RMS}\sim4$~nJy, this leads to a dynamic range requirement of $D\ga 4\times10^{7} = 76$~dB.  
Reaching this dynamic range will likely require, at minimum, antennas having extremely stable beams and pointing capabilities. 
Such high precision antennas should function well at frequencies $\ga$10~GHz, which is the current frequency cutoff of the SKA Phase-1 and Phase-2.  

Table \ref{tbl-2} summarizes the requirements on the SKA in order that it
be able to detect galaxies forming stars with rates $\ga$25~$M_{\sun}~{\rm yr}^{-1}$ at all redshifts where such galaxies may be expected.  
As such, the SKA would therefore be sensitive to a significant number of the high-$z$ galaxies reported in the UV luminosity function studies of \citet{rb07,rb08}.   
However, there are still large uncertainties in the current understanding of the sub-$\mu$Jy population of radio sources, and achieving such a high imaging dynamic range represents a significant technical challenge.  
The current generation of SKA pathfinders (e.g. \hbox{EVLA}, \hbox{ASKAP}, \hbox{MeerKAT}, and \hbox{APERTIF}) could usefully reduce the uncertainties in both the sub-$\mu$Jy radio source population(s) and high dynamic range imaging.

Finally, as the observed radio continuum emission from normal star-forming galaxies at high redshift  will likely be dominated by their thermal emission, this analysis does not require that dust be present in these galaxies for these sensitivity calculations to be valid.   
Specifically, this analysis requires their bolometric luminosities, and not necessarily their IR luminosities, be $\ga10^{11}~L_{\sun}$ for detection.  
Thus, radio continuum emission from galaxies becomes the only way to find and quantify the star formation properties of galaxies unbiased by dust.  
The importance of this fact for studies of the cosmic star formation history are discussed in $\S$\ref{sec-sfr}.  

\begin{deluxetable}{lc}
\tablecaption{SKA Deep Continuum Survey Assumptions \label{tbl-2}}
\tablewidth{0pt}
\tablehead{
\colhead{Parameter} & \colhead{Value}
}
\startdata
Target Star Formation Rate          & 25~$M_\sun$~yr${}^{-1}$ \\
\hline
Bandwidth (BW) & 1~GHz\\
Angular Resolution ($\theta_{\rm S}$) & $\la0\farcs3-1$\arcsec\\
Primary Beam ($\Omega_{\rm b}$) &  $\approx 1$~deg$^{2}$\\ 
Imaging Dynamic Range ($D$) & $\ga 76$~dB\\
Integration Time ($t_{\rm int}$) &300~hr ($\sim$1~Ms)\\
\hline
SKA Sensitivity ($A_{\rm eff}/T_{\rm sys}$) & $15000~{\rm m^{2}~K^{-1}}$\\
Image RMS Noise Level ($\sigma_{\rm RMS}$) & $\approx$4~nJy
\enddata
\end{deluxetable}

\section{Discussion}
I will now discuss how various galaxy properties can be inferred from discrepancies between expected and observed IR/radio ratios.  
I will also discuss what this means for the prospect of using deep radio continuum surveys, both at high and low frequencies, to study star formation over cosmic time using the SKA.  

\subsection{Searching for the Emergence of Large-Scale Magnetic Fields}
One of the primary energetic constituents of the ISM is its magnetic field.   
Its energy density is equal to that of a galaxy's radiation field, CRs, and turbulent gas motions, yet its origin and role in shaping galaxy evolution, in particular through the possible regulation or star formation, remains highly uncertain  \citep[e.g.][]{kf01,cox05}.  
At present the leading theory for large-scale magnetic field generation is the mean-field dynamo model \citep[e.g.][]{rb96}.  
In this paradigm galaxy-scale magnetic fields are built up over time, suggesting that field strengths should be typically much weaker in the past.  
The exact details on how the amplification of weak, seed magnetic fields in the halos of protogalaxies eventually translates into galaxy-scale fields is uncertain.  
Does the first epoch of star formation quickly initiate a dynamo in a galaxy?  
Does the field form during the in-fall of gas prior to star formation?  
Recent theoretical work suggests that weak, seed magnetic fields may be quickly amplified by the turbulent (small-scale) dynamo, yielding strong ($\sim$20~$\mu$G) turbulent fields by $z\sim 10$ \citep{ta09}.  
These fields then serve as a seed for the mean-field (large-scale) dynamo, resulting in regular fields of $\mu$G strengths having a few kpc coherence within $\sim$2~Gyr (i.e. by $z\sim3$).  
Such a scenario is consistent with indirect evidence from Faraday rotation studies of distant radio sources which claim $\mu$G fields to be in place for normal galaxies at $z\sim1.5$ \citep{pk08,mb08}.  


Since the non-thermal radio emission from galaxies, and therefore the linearity of the FIR-radio correlation, depends critically on the magnetic field strength of a galaxy, deviations from nominal ratios as a function of redshift may help to illuminate the presence and characteristics of magnetic fields in galaxies at early epochs.  
However, given that regular magnetic fields are thought to weaker in the past, as suggested by the mean-field dynamo model, maintaining a constant IR/radio ratio with increasing redshift by increasing magnetic field strength with redshift is somewhat contrived.   
If the FIR-radio correlation is found to hold for normal galaxies at $z\ga$3, either the the turbulent dynamo and associated small-scale magnetic fields play an important role at early epochs, or our physical picture for the correlation needs revision.
This assumes of course that the existence of an AGN in these galaxies, whose radio emission may fortuitously bring the IR/radio ratio to near or below the local correlation value, can be ruled out.
Such situations are discussed below in $\S$\ref{sec-data} where I compare observations with these expectations.    

In the absence of strong magnetic fields at high-$z$, the IR/radio ratios of galaxies should be large, similar to what is observed for `nascent' starbursts.    
These systems are rare ($\la$1\%) in the local Universe \citep{hr03}, and are thought to be experiencing the onset of an episode of star formation after a long ($\approx$10$^{8}$~yr) period of quiescence.  
Since OB stars will always ionize the nearby gas, generating free-free emission, these galaxies may only be identifiable in the radio by their thermal radio continuum emission. 



\begin{figure*}
\plotone{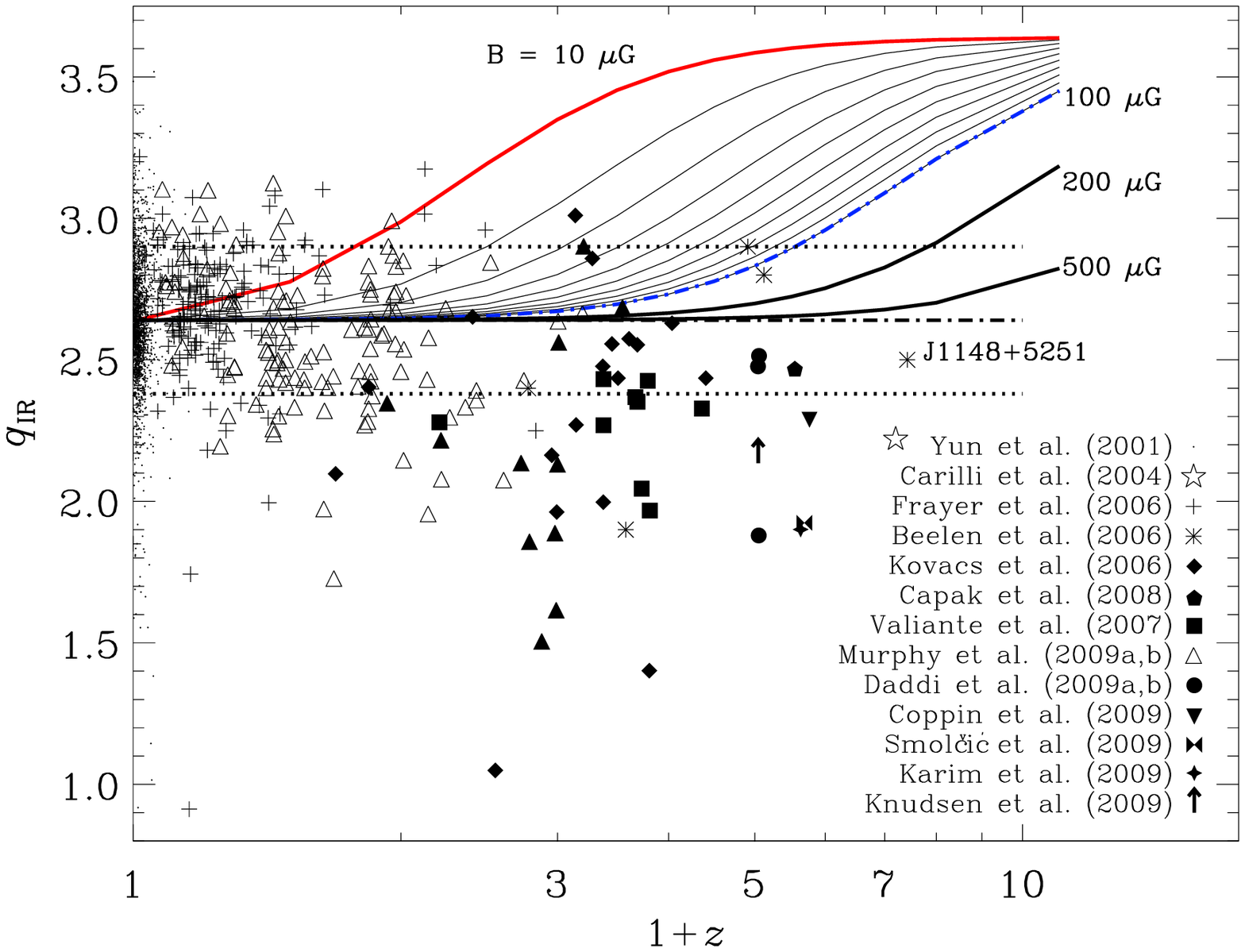}
\caption{
The same as Figure \ref{fig-5}, except that I have plotted actual data from a number of studies in the literature.  
A logarithmic redshift scale is used since there are so few points at $z\ga3$.  
Filled in symbols indicate that an object is an SMG, i.e. the entire \citet{ak06}, \citet{ev07},\citet{pc08},  \citet{ed09a,ed09b}, \citet{kc09}, \citet{vs09}, \citet{ak09}, and \citet{kk09}  samples, as well as half of the  \citet{ejm09a} sample.  
The lower limit ($uppward$-$arrow$) for the \citet{kk09} data point is due to only having an upper limit at 1.4~GHz.  
The radio-quiet quasar J114816+525150 is labeled; at a redshift of $z=6.64$ it lies at $\sim$1/16 of the Universe's current age.  
Rather than seeing galaxies tend toward higher IR/radio ratios per expectations given that IC losses of the CMB become increasingly important with increasing redshift, many galaxies tend to have IR/radio ratios lower than the local value (i.e the SMGs).  
\label{fig-6}}
\end{figure*}

\subsection{Current Observational Results \label{sec-data}}
Using existing facilities a number of deep FIR, submm, and mm surveys have detected galaxies at fairly high redshfits.  
Sources detected in the FIR and radio from a number of these studies are included in Figure \ref{fig-6} with the highest redshift sources being the $z\sim6$ quasars detected by \citet{cc04} and \citet{ab06}.   
In cases where only a FIR measurement is provided, the assumption that ${\rm IR(8-1000~\micron) \approx 2\times FIR(42-122~\micron)}$ is made, which is exactly the difference between the average FIR/radio and IR/radio ratios reported by \citet{yrc01} and \citet{efb03}, respectively.   
The average and $\pm~1\sigma$ lines for the local IR-radio correlation are drawn \citep[i.e. $q_{\rm IR} \approx 2.64 \pm 0.26$~dex;][]{efb03}.  
As in Figure \ref{fig-5}, the expected locations of a typical star forming galaxy as IC losses of the CMB become increasingly important relative to the internal synchrotron losses of CR electrons for various magnetic field strengths are overplotted.  
Since so few sources are detected at $z\ga3$, a logarithmic scaling is used for the redhift axis to better identify any trends.  

\subsubsection{Submillimeter Galaxies}
Most of the galaxies out to $z\sim1$ \citep[e.g. the 70~$\micron$ detected sources in; ][]{df06,ejm09b} tend to have IR/radio ratios similar to the local value, which is consistent with the expectations presented here since the magnetic fields in such galaxies need not be excessively large to compete with IC losses off of the CMB.  
Once beyond $z\sim1$ there is an increasingly larger number of galaxies having IR/radio ratios below the canonical value, rather than larger per the suggested expectations.  
The majority of these sources are submillimeter galaxies (SMGs).  
While typically detected in the hard-band ($2-8$~keV) X-rays,  thus indicating the presence of an AGN, this result is somewhat surprising since SMGs are thought to be primarily powered by star formation \citep[e.g.][]{da05, ejm09a}. 
 
The samples of \citet{ak06} and \citet{ev07} consist entirely of SMGs and the filled triangles indicate the SMGs included in \citet{ejm09a}.  
The mean redshift among these three samples of SMGs is $z \approx2.17\pm0.65$.  
\citet{ak06} report IR/radio ratios for their sample of SMGs to be systematically below the local value.  
\citet{ejm09a} also found  that the SMGs in their sample to have IR/radio ratios which are, on average, a factor of $\sim$3.5 lower than the canonical ratio.  
The mean IR/radio ratio for the sample of \citet{ev07} is $\sim2.27\pm0.16$~dex, a factor of $\sim$2 below the local value of 2.64~dex. 
The average IR/radio ratio among all 38 of these $z\sim$2.2 SMGs is $q_{\rm IR} = 2.26\pm0.41$~dex, a factor of $\sim$2.4 times smaller that the local value.  
A number of possible explanations for why these systems depart for the FIR-radio correlation are included in \citet{ejm09a}, but a definitive conclusion has not been drawn.  
 
Recently, a number of higher redshift SMGs have been identified \citep[][]{pc08,ed09a,ed09b,kc09,vs09,ak09,kk09}.  
The average redshift among these 8 objects is $z \approx4.40\pm0.33$.  
Similar to the $z\sim2.2$ SMGs, the mean IR/radio ratio of these $z\sim4.4$ SMGs is $q_{\rm IR} = 2.20 \pm 0.27$~dex, a factor of $\sim$2.8 times smaller that the local value. 
This result is somewhat surprising since the expected increase in the IR/radio ratios between a $z\sim2.2$ and $z\sim4.4$ galaxy due to the increase in IC losses to CR electrons from the CMB is a factor of $\sim$3 for a $\approx$35~$\mu$G field, the average value for the \citet{ejm09a} SMGs using the revised minimum energy formula of \citet{bk05}.  

While it has been suggested that the properties of the $z\sim4.4$ SMGs are similar to the  $z\sim2.2$ SMGs \citep[e.g.][]{ed09a}, the consistency in their IR/radio ratios argues the opposite, suggesting that their star formation driven IR and/or radio continuum emitting properties are sufficiently different to account for the large discrepancy arising from extra cooling of CR electrons by IC scattering off the CMB.  
Alternatively, this result may also simply suggest that, unlike their FIR output, the radio emission from both populations of SMGs is in fact dominated by AGN, or that their magnetic fields are significantly strong (i.e. $\ga 300~\mu$G) such that $U_{B} \gg U_{\rm CMB}$ at these redshifts.  
The latter explanation is tough to reconcile since, in the presence of such strong fields, CR electrons emitting at GHz frequencies will be cooled primarily by ionization and bremsstrahlung losses rather than by synchrotron radiation (see $\S$2.2), unless some other physics is at work.      
At present it is difficult to make any strong conclusions from this comparison since it suffers from low number statistics; very few SMGs that have been identified at $z\ga4$.    


\subsubsection{Radio Quiet QSOs at High-z}
The samples of \citet{cc04} and \citet{ab06} consist entirely of radio-quiet quasars QSOs, a total of 6.  
The highest redshift QSO, J114816+525150 (hereafter, J1148+5251), is identified in Figure \ref{fig-6} and appears to have an IR/radio ratio which is amazingly consistent with the local value.  
In fact, 4/6 (i.e. $\sim$66\%) of the QSOs have IR/radio ratios that are within 1$\sigma$ of the canonical ratio.  
While these observations are suggestive that the energetics of these objects is dominated by star formation, such a conclusion is not so clear.  

These galaxies may lie near the nominal IR/radio ratio due to excess emission from their AGN.  
Even if all of the synchrotron emission associated with ongoing star formation in these galaxies is suppressed due to other CR electron cooling processes competing successfully with respect to synchrotron radiation, the excess emission from an AGN may be enough to fortuitously put a galaxy back on the local correlation.  
This scenario is likely the simplest way to explain how QSO's at these redshifts have IR/radio ratios consistent with normal star-forming galaxies in the local Universe.  

Alternatively, these objects would require magnetic field strengths nearing $\ga$500~$\mu$G levels to remain on the local FIR-radio correlation if powered by star formation alone.  
Taking J1148+5251 as an example; the galaxy is being observed $\la$1~Gyr after the Big Bang.  
While it has been suggested that strong ($\sim$20$~\mu$G) turbulent fields may be present by $z\sim10$, it is unclear how these seed fields could be amplified so rapidly and form galaxy-wide large-scale regular fields of $\ga200-500~\mu$G.  
If global star formation initiates a large-scale dynamo which acts to regulate the magnetic fields of galaxies \citep[e.g.][]{my73,mr87,ruz88}, the timescale for this must occur rapidly as star formation is only $\sim$400~Myr old assuming that star formation began around $z\sim10$.  
The requirement for such a rapid amplification phase suggests that the large-scale field may begin to form before the initiation of star formation, during the in-fall of gas and dust \citep[e.g.][]{ps89,rb94,hb93,kuls97}.  
However, these strong magnetic fields would then have to decay considerably over cosmic time to recover the typical strengths observed for galaxies in the local Universe.  
Detections of less exotic systems at these redshifts in the FIR and radio (e.g. using ALMA and the SKA) will be necessary to reconcile these outstanding questions.  
In fact, a comparison of the IR/radio ratios for these high-$z$ galaxies will likely be one of the only methods to test for the existence and strength of magnetic fields in the early Universe while trying to assess what role they may play in star formation.  

\subsubsection{Lyman Break Galaxies at $z\sim3$}
\citet{cc08} recently analyzed the radio properties for a large sample of Lyman break galaxies (LBGs) at $z\sim3$.  
Using a stacking analysis, and assuming the local FIR-radio correlation, they find that the radio (1.4~GHz) derived star formation rate is a factor of $1.8\pm0.4$ times larger than that from observed UV measurements.  
Assuming that this difference arises from attenuation by dust at UV wavelengths, this value is significantly less than the factor of $\sim$5 normally assumed for LBGs.  
They suggest that this discrepancy may be the result of depressed radio emission resulting from increased IC cooling of CR electrons off of the CMB, consistent with the analysis presented here.  

The stacked 1.4~GHz flux density reported by \citet{cc08} is $0.90\pm0.25~\mu$Jy, which corresponds to a rest-frame 1.4~GHz flux density of 0.68~$\mu$Jy assuming a radio spectral index of $0.8$ \citep{jc92}.  
Taking the typical size of an LBG to be $\la$1$\arcsec$ \citep[e.g.][]{mg02b} results in a minimum energy magnetic field of $\ga 8~\mu$G \citep{bk05}.  
The minimum energy magnetic field is relatively insensitive to the radio surface brightness (i.e. $B_{\rm min} \propto I_{\nu}^{1/(3+\alpha)}$); 
even if the reported 1.4~GHz flux density were under-predicted by a factor of 10, the corresponding magnetic field strength would only increase by a factor of $\approx$1.8.
The increased energy loss to CR electrons from IC scattering off of the CMB at $z\sim 3$, assuming equipartition between the galaxy's magnetic and radiation field energy densities, will decrease the observed 1.4~GHz flux density by a factor of $\sim$3.5 which is easily able to account for the discrepancy between dust attenuation factor suggested by the radio (1.4~GHz) derived star formation rate of \citet{cc08} and the factor of $\sim$5 normally assumed for LBGs.  
If the intrinsic magnetic fields of LBGs are as large as the minimum energy fields for SMGs (i.e. $\sim 35~\mu$G,) then the observed non-thermal radio continuum emission should still be suppressed by a factor of $\sim$2.  

\subsection{Implications for Radio-Based Studies of Star Formation at High Redshifts \label{sec-sfr}}
While I have looked to see how the GHz emission from galaxies will be affected by increased CR electron cooling as the result of IC scattering off of the CMB at high redshifts, I  have yet to look at what this may do for higher frequency radio observations.  
I now consider the potential usefulness of low ($\la$2~GHz) and high ($\ga$10~GHz) frequency radio observations for measuring star formation over cosmic time.  

\subsubsection{At Low Frequencies: $\la$2~GHz}
As discussed above, studies of star formation at low frequencies using the SKA will be complicated by the expected evolution in the FIR-radio correlation with increasing redshift.  
When coupled with the additional emission from AGN, which may appear to place galaxies back on the FIR-radio correlation, it appears that using GHz observations to estimate the cosmic star formation history may lead to increasingly unreliable estimates of star formation rates with increasing redshift.   
Galaxies which host compact starbursts may have large magnetic fields and gas densities in which case electrons contributing to GHz emission will lose their energy primarily to ionization and bremsstrahlung processes rather than synchrotron radiation.  
These galaxies may move off the correlation by some fixed amount depending on how the various energy-loss terms compete, resulting in incorrect radio-based estimates of star formation rates using locally-derived calibrations.  
In conclusion, it seems that interpreting the non-thermal GHz measurements of high-$z$ galaxies properly may prove difficult.  

\begin{figure}
\plotone{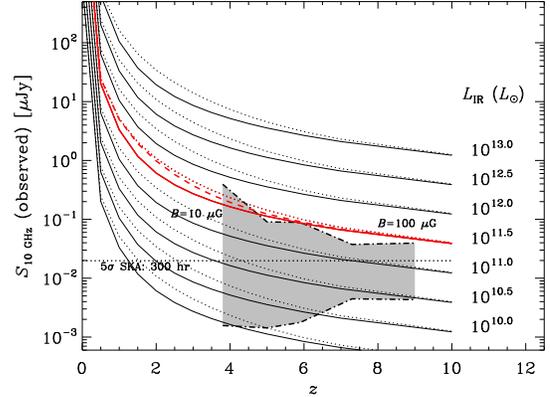}
\caption{
The same as Figure \ref{fig-3}, except that the expected emission at 10~GHz is considered. 
The expected sensitivity of the SKA at 10~GHz is also assumed identical to the 1.4~GHz sensitivity.  
Since the non-thermal contribution is almost completely suppressed at high redshift,  the 10 and 1.4~GHz emission should be detectable from the same population of galaxies at these epochs.  
The expected flux density range for galaxies included in the $z\sim$ 4, 5, 6, 7, and 9 UV luminosity function studies of \citet{rb07,rb08} is again given by the shaded region.  
\label{fig-7}}
\end{figure}

\begin{figure}
\plotone{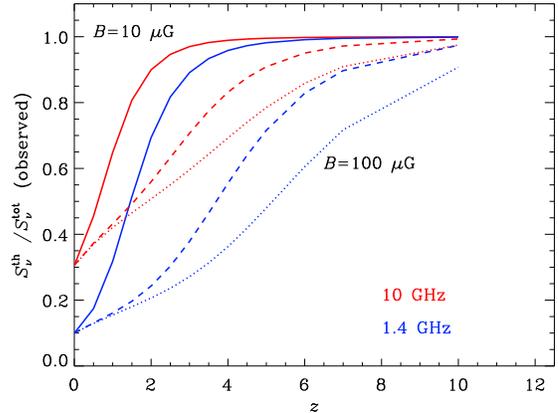}
\caption{
The expected thermal fraction for observed-frame 1.4 (blue) and 10~GHz (red) observations of normal star-forming galaxies vs. redshift.  
Intrinsic magnetic fields of 10 ({\it solid}-lines), 50 ({\it dashed}-lines), and 100~$\mu$G ({\it dotted}-lines) are shown.  
By observing at 10~GHz, the observed radio emission is dominated by thermal processes, making it an ideal measure for the massive star formation rate of galaxies, beyond a redshift of $z\sim2$ even for magnetic field strengths of $\sim$100~$\mu$G.  
\label{fig-8}}
\end{figure}

\subsubsection{At High Frequencies: $\ga$10~GHz}
While deep continuum studies aimed at measuring the star formation properties of high-$z$ galaxies at low frequencies (i.e. $\la$2~GHz) are not straightforward due to large variations in the amount of non-thermal radio emission, such studies at slightly higher frequencies (i.e. $\ga$10~GHz) are likely to be much more promising.  
In fact, it can be argued that deep radio continuum studies at these frequencies will provide the most accurate estimates for the star formation rate of high-$z$ galaxies.  
So long as average galaxy magnetic fields do not exceed mG strengths, which seems unlikely given that at such high redshifts it is unclear how such strong fields could be seeded, the observed radio emission at $\ga$10~GHz should be dominated by thermal (free-free) radiation directly associated with Lyman continuum photons emitted by newly formed massive stars.  

In Figure \ref{fig-7} the expected observed-frame 10~GHz continuum flux densities of galaxies are plotted for a range of IR luminosities under the same assumptions that went into creating Figure \ref{fig-3} (see Table \ref{tbl-2}).  
Overplotted is the expected range in observed-frame 10~GHz flux density for the sources studied by \citet{rb07,rb08}.  
It is shown that even at 10~GHz the same population of LIRGs, if they exist, should be detectable at all redshifts.  
Furthermore, it appears that the internal magnetic strength of a galaxy does not play as signifiant a role in a source's detectability since the thermal fraction at these frequencies is so large, and becomes increasingly so with redshift.  
This is illustrated in Figure \ref{fig-8} where the expected thermal fraction for observed 1.4 and 10~GHz flux densities is plotted as a function of redshift for galaxies having 10, 50 and 100~$\mu$G fields.  

Observations at 10~GHz are dominated by thermal emission for $B\la100~\mu$G beyond a redshift of $\sim$2.  
The same cannot be said for 1.4~GHz observations until the galaxy population being observed lies at a redshift of $z\ga$5.
At 10~GHz, the radio emission from $z\ga5$ galaxies is likely to be $\ga$80\% thermal emission.  
Therefore, it is believed that deep continuum surveys at $\ga$10~GHz may provide an excellent measure for the star formation activity in the highest redshift galaxies.  

This conclusion will only be valid if there is a negligible amount of microwave emission associated with dust.  
While thermal emission from interstellar dust grains is negligible between $10-100$~GHz, so called 'anomalous' dust emission, possibly arising from a population of ultrasmall, rapidly rotating grains with a non-zero electric dipole moment, may contribute significantly at microwave wavelengths \citep[e.g. ][]{dl98a,dl98b,ahd09}.  
If significant, high-frequency radio observations would likely overestimate true star-formation rates.  
This occurrence would also require that the observed high-$z$ galaxies have a significant amount of ultrasmall dust grains.  

Comparing these observations with results from ALMA, which should be able to detect the peak FIR continuum emission from galaxies at similar high redshifts \citep[i.e. $z\ga$5;][]{am01}, should help disentangle this issue and result in powerful star formation and AGN diagnostics.  
Such tools will be critical for aiding in the interpretation of deep SKA imaging surveys which, unlike ALMA, will be able to survey large areas of the sky relatively quickly.   
And, as previously stated, even if the dust content in typical galaxies at these early epochs is negligible such that their FIR continuum emission is invisible to ALMA, radio continuum emission will likely be the best means to identify such objects and and quantify their star formation properties. 

\subsection{Other Physical Considerations}
In the above presentation I have only considered variations to the observed IR/radio ratios which arise from an increase in the competition of energy-loss processes to primary CR electrons relative to synchrotron radiation.  
Alternatively, other physical processes may have their hand in driving systematic variations in a galaxy's IR/radio ratio, perhaps even keeping the correlation linear versus redshift fortuitously.  

\subsubsection{Implications for the Possible Size Evolution of Galaxies}
It is possible that the suppression of a galaxy's non-thermal emission may not be as significant as previously suggested if the size of galaxies decrease with increasing redshift, resulting in larger values of $U_{B}$ per unit star formation.   
For instance, \citet{rb04} suggest that the half-light radius of galaxies between $2 \la z \la 6$ follow an $r_{\rm hl}\sim(1+z)^{-1}$ relationship.  
If true, the $U_{\rm CMB} \propto (1+z)^{4}$ quenching of galaxy synchrotron emission is effectively reduced by a factor of $(1+z)^{2}$ for a constant magnetic field strength, and IC losses off the CMB will not be as significant for CR electrons.  
In this case such a strong departure from the canonical FIR-radio correlation may not be as easily seen.  
However, if galaxies do become more compact, the escape times for CR electrons to leave the galaxy will likely decrease resulting in a decrease to the observed synchrotron emission.    
And, if the ISM of such galaxies are typically clumpy, similar to local dwarf irregular galaxies, then escape times should decrease even more \citep[e.g.][]{ejm08}.  

For a random walk diffusion, the escape time is roughly proportional to $l_{\rm esc}^{2}$.  
Assuming that $l_{\rm esc} \sim r_{\rm hl} \sim(1+z)^{-1}$, the escape time will decrease as $(1+z)^{-2}$ which should compensate for the increase to $U_{B}$ by that same factor.    
Furthermore, if these compact galaxies also host starbursts, advection out of each system due to a galactic wind may lead to even shorter escape times resulting in even larger IR/radio ratios at these early epochs.  
Therefore, even if galaxies are smaller at high-$z$, evolution in the FIR-radio correlation with redshift is still expected to be observed such that IR/radio should be larger with increasing $z$.  
Given that local starbursting galaxies follow the FIR-radio correlation, the suggestion for increased losses due to advection by galactic winds in such high-$z$ systems would imply that they are different than starbursts observed locally.   

\subsubsection{Early Chemical Enrichment and the First Dust Grains \label{sec-dust}}
The FIR-radio correlation may also fail at high redshifts for a variety of reasons arising from differences in a galaxy's dust emission properties.  
For instance, IR/radio ratios should depart from the nominal value if dust is not present, the size distribution of grains take on a different form,  or if the FIR emitting properties of dust grains are significantly different than in the local Universe.  
Currently, observations of quasars and luminous galaxies at high-$z$ show that these systems contain large masses of dust \citep[e.g.][]{wang08}.  
While it is unlikely that these dusty systems are the representative population at early epochs, it is worth noting that the dust contained in these very young systems does not appear to be exotic enough to move systems off of the FIR-radio correlation.    
For example, the $z=6.42$ quasar (J1148+5251), which is plotted in Figure \ref{fig-6} using data taken from \citet{ab06}, has an estimated dust mass ranging between $2\times10^{8}~M_{\sun}$ \citep{ed07} and  $7\times10^{8}~M_{\sun}$ \citep{fb03}.  
Dust-to-gas mass ratios are also found to be $\ga$0.004 which is similar to, or exceeding the, $\sim$0.005 value of the Galaxy.  

Using this QSO as an example, \citet{bd09} states that SNe are likely responsible for producing both the metals which compose grains, as well as supernovae-condensed dust to provide surface areas for grain growth.  
However, \citet{bd09} also shows that the dust in these objects are not predominantly composed of SN dust, known to have different emitting properties than that of typical dust grains in normal star-forming galaxies.  
Rather,  the bulk of the dust formed in these high-$z$ galaxies is attributed to grain growth occurring more rapidly than grain destruction in the ISM, resulting in large dust masses that are already in place in galaxies 400~Myr after the first episode of star formation.  
This estimate assumes that star formation within galaxies began at $z=10$.  
Given that the dust-to-gas ratios in these high-$z$ galaxies do not appear significantly different than the Milky Way value, and that the grain formation process (i.e. {\it in~situ}) is similar to that of normal star-forming galaxies in the local Universe \citep{bd09}, it seems reasonable to believe that the FIR dust emission from these galaxies does not drive large variations in their IR/radio ratios relative to the expected variations in their radio continuum emission.     

\subsubsection{Effect of CR Electron Acceleration Efficiencies \label{sec-acc}}
The efficiency in which CRs are accelerated in SNRs plays a significant role in determining the synchrotron emissivity from galaxies.  
Furthermore, the distributed re-acceleration of CR electrons arising from external interactions such as ram pressure may add significant flux to the observed radio continuum from galaxies \citep[e.g.][]{ejm09c}.  
The efficiency with which the initial energy of the SNe explosion is converted into kinetic energy depends on the relative importance of radiation and adiabatic losses.  
In systems having extremely dense ISM, the adiabatic phase of the supernovae expansion may be halted as it expands into the ambient medium, thus reducing the amount of energy lost to adiabatic expansion.  
It is possible that this extra energy could be, at least partially, used in the acceleration of CRs yielding a larger amount of synchrotron emission per unit star formation than in normal galaxies.  
For the SMGs, whose IR/radio ratios are systematically below the FIR-radio correlation by $\sim2-3$ (see $\S$\ref{sec-data}), this scenario does not appear to help explain their low ratios.  

As an example, the modeling of \citet{ed91,ed00} shows that the total acceleration efficiency for CRs increases from $\sim$0.15$E_{\rm SN}$ to $\sim$0.25$E_{\rm SN}$, where $E_{\rm SN}=10^{51}$~erg is the total explosion energy of the SNe, when the external ISM density is increased from $\sim$1~cm$^{-3}$ to $\sim$10~cm$^{-3}$.  
The ISM densities of starbursts are likely much larger, as with the case of the example where $n_{\rm ISM} \sim 10^{4}$~cm$^{-3}$ was assumed, thus the evolution of SNRs will likely be different and may work to increase the synchrotron emissivity in such galaxies through an increased efficiency in particle acceleration.  
However, unless SMGs host starbursts that are sufficiently different than those in local starbursting galaxies, which do not appear to exhibit such variations in their IR/radio ratios, this scenario alone might not be sufficient to explain their low IR/radio ratios.  
At present, more simulation work on this problem is needed.    

\subsubsection{Secondary Electrons and Positrons \label{sec-sec}}
So far synchrotron radiation arising for primary CR electrons accelerated in SNRs has been considered.  
CR nuclei, which are accelerated in the same SNRs, will inelastically scatter off of the interstellar gas yielding $\pi^{0}$ particles that decay into secondary $e^{\pm}$'s and $\gamma$-rays, and $\pi^{\pm}$ particles which will decay into secondary $e^{\pm}$'s and neutrinos.  
These secondary $e^{\pm}$'s will undergo the same energy-loss processes described in $\S$2.1 and $\S$2.2.  
At $\sim$GeV energies the ratio of secondary/primary electrons in the Galaxy is roughly 2:1, and decreases to nearly 1:3 at $\sim$3~GeV energies \citep{tp08}.  
The large secondary fraction arises from the fact that there are many more CR nuclei than electrons.  
Consequently, diffusion models suggest that $\sim$20-50\% of the diffuse GHz synchrotron emission in the Galaxy arises from secondaries.   
However, secondary particles are produced more diffusively than primary electrons, thus their synchrotron morphology should not show any coincidence with their sites of origin.  
Given that the FIR and radio images of galaxies are remarkably similar \citep[e.g.][]{ejm06,ejm08} argues that the secondaries likely contribute mostly near or around star-forming regions.  

In the case of a starburst, having a much larger ISM density, the cross-section for collisions between CR nuclei and the interstellar gas is increased, thus increasing the number of $e^{\pm}$'s for a fixed primary nuclei/electron ratio which may actually dominate the diffuse synchrotron emission.    
Determining the contribution of additional synchrotron emission arising by an increased population of secondary $e^{\pm}$'s clearly goes beyond the scope of this paper, however the fact that this effect may play a role is noteworthy.    
The recently launched $Fermi$ $\gamma$-ray observatory may help to shed some light on this topic.  

\subsubsection{Variations in the Initial Mass Function \label{sec-imf}}
While star-forming galaxies are currently thought to be responsible for completely reionizing the intergalactic medium (IGM) by $z\sim6$ \citep[e.g.][]{rb01,xf06}, the ionizing flux arising from star formation in Lyman break galaxies at similar redshifts \citep{rb07,rb08} appears to fall a factor of $\ga6$ below the minimum value required to maintain an ionized IGM for a given clumping factor and escape fraction under the assumption of a Salpeter stellar initial mass function (IMF).  
\citet{rrc08} has shown that this discrepancy can be reconciled by flattening the stellar IMF to have a slope of $\sim-1.7$ if reionization occurred at $z=9$.  
The idea of a top heavy IMF at these epochs does not seem completely inappropriate given that low metallicity environments will favor the production of more high-mass stars.  
In this case, one expects there to be both more SNe, increasing the amount of non-thermal synchrotron emission, as well as an increase to the ionizing photon rate per baryon, leading to an increase in the thermal radio continuum emission.  
It is worth noting that only a few SNe are required to enrich a galaxy from primordial values to a few tenths solar metallicity, thus most galaxies observed with ages larger than a dynamical timescale are not likely to be that metal poor \citep[e.g.][]{rw09,hy09}.  

Using the results of \citet{rrc08}, the ionizing photon rate for a stellar IMF having a slope of $-1.7$ is a factor of $\approx$1.7 times larger than a that for a typical Salpeter IMF slope of $-2.3$ over a mass range of $1-200~M_{\sun}$.  
Similarly, for a given star formation rate, such a change to the slope of the stellar IMF should result in an increase to the core collapse supernova rate by a factor of $\approx$2 \citep[e.g.][]{pm98}.  
Thus, if the stellar IMF in these high redshift systems transitions to being top heavy, the results presented here should not be significantly affected as both the thermal and non-thermal radio continuum output should change by factors of $\la$2.     
In the case that there is a significant amount of dust in these high-$z$ systems, and assuming that their non-thermal radio continuum emission is completely suppressed by the additional IC losses to CR electrons by the CMB, their IR/radio ratios should still remain similar for a varying IMF given that both the FIR and thermal radio emission are directly proportional to the ionizing photon rate.  
However, since the increase to the ionizing photon rate will translate into an increase to the thermal radio emission, careful analyses of radio source counts at these redshifts, along with comparison between rest-frame UV studies, may help to illuminate possible variations to the stellar IMF.  } 

\section{Summary}
In this article I have put together a predictive analysis for the expected evolution of the FIR-radio correlation versus redshift arising from variations in the CR electron cooling time-scales as IC scattering off of the CMB becomes increasingly important.   
As a result, I have looked at how discrepancies between observed and predicted IR/radio ratios can be used to characterize the magnetic field properties of galaxies at high-$z$, where the application of other observational indicators are extremely difficult.     
In doing so, I have also looked at the value of deep radio continuum surveys in studies of star formation at high-$z$, particularly in the context of the SKA.  
My findings can be summarized as follows:

\begin{enumerate}

\item Deep radio continuum observations at frequencies $\ga$10~GHz using next generation facilities like the SKA will likely provide the most accurate measurements for the star formation rates of normal galaxies at high~$z$.   
The non-thermal emission from such galaxies should be completely suppressed due to the increased IC scattering off of the CMB leaving only the thermal (free-free) emission detectable; 
this situation may be complicated if `anomalous' microwave emission from spinning dust grains \citep[e.g.][]{dl98a} in such systems is not negligible.  

\item For normal star-forming galaxies to remain on the local FIR-radio correlation at high redshifts requires extraordinarily large magnetic field strengths to counter IC losses from the CMB.  
For example, the magnetic field of a $z\sim6$ galaxy must be $\ga$500~$\mu$G to obtain a nominal IR/radio ratio.  
Thus, galaxies which continue to lie on the FIR-radio correlation at high-$z$, such as the sample of radio-quiet QSO's, most likely have radio output which is dominated by an AGN, or the physics at work is simply unknown.  

\item The average IR/radio ratio among a sample of $z\sim2.2$ SMGs is consistent with that for a sample of $z\sim4.4$ SMGs.  
This is in strike contrast with the factor of $\sim$3 increase expected due to the increase in energy losses to CR electrons from IC scattering off of the CMB between these two redshifts.
While this comparison suffers from low number statistics, it suggests that either the star formation driven IR and/or radio continuum emitting properties of these two populations of SMGs are different, or that AGN dominate the radio output of both of them.  
Simply assuming that their magnetic field strengths are significantly large (i.e. $\ga 300~\mu$G) is problematic since, in the presence of such strong fields, CR electrons emitting at GHz frequencies will cool primarily by ionization and bremsstrahlung processes rather than by synchrotron radiation which should lead to large IR/radio ratios unless other physics is at work.


\item To detect typical LIRGs ($L_{\rm IR} \ga 10^{11}~L_{\sun}$) at all redshifts will require nJy sensitivities at GHz frequencies, specifically a 5$\sigma$ point-source sensitivity of $\approx$20~nJy (i.e. $\sigma_{\rm RMS} \approx 4$~nJy).  
Thus, for the SKA to achieve this sensitivity for a reasonably long (300~hr) integration necessitates that $A_{\rm eff}/T_{\rm sys} \sim 15000$~m$^{2}$~K$^{-1}$.  
At this sensitivity, the SKA will be sensitive to all galaxies forming stars at $\sim$25~$M_{\sun}~{\rm yr}^{-1}$.  
This includes a significant amount of sources included in the UV luminosity function work of \citet{rb07,rb08} between $4\la z \la 9$.  

\item Deviations in the IR/radio ratios of normal star-forming galaxies at high redshifts will likely provide one of the only ways to put constraints on the existence and strength of magnetic fields in the early Universe.  
For instance, if one knows the radio spectral index, galaxies with magnetic fields $\la$50~$\mu$G should be easily identifiable once beyond a redshift of $\ga$4 as their IR/radio ratios should be a factor of $\ga$4 larger than the nominal value, on average.   

\end{enumerate}

\acknowledgments
I would like to thank Joe Lazio for initially suggesting this study, as well as the rest of the organizers of "The Community Workshop on High-Dynamic Range Imaging Continuum" held in Cape Town, South Africa during February, 2009.
I am also extremely grateful to George Helou, Jim Condon, Rainer Beck, Ranga-Ram Chary, Troy Porter, and Brian Lacki for stimulating discussions which greatly helped improve the quality of the paper.

\end{document}